\documentclass[amsmath,amssymb,floatfix,aps,showpacs,superscriptaddress,nofootinbib,notitlepage,twocolumn,pra]{revtex4-1}

\usepackage{amsmath,amsthm,amssymb,bm,hyperref,graphicx,color,float,changes,mathtools,enumitem,tikz}
\allowdisplaybreaks

\newtheorem{prop}{Proposition}

\newtheorem{thm}{Theorem}[section]

\newcommand{\be}{\begin{equation}}
\newcommand{\ee}{\end{equation}}
\newcommand{\bea}{\begin{eqnarray}}
\newcommand{\eea}{\end{eqnarray}}
\newcommand{\6}{\partial}

\newcommand{\inti}{\int_{-\infty}^{+\infty}}

\newcommand{\Gm}[1]{\mathcal{G}_{#1}^{\mbox{\scriptsize{(--)}}}}
\newcommand{\Gp}[1]{\mathcal{G}_{#1}^{\mbox{\scriptsize{(+)}}}}

\begin{document}

\title{Correlation functions of one-dimensional strongly interacting two-component gases}

\author{Ovidiu I. P\^{a}\c{t}u}
\affiliation{Institute for Space Sciences, Bucharest-M\u{a}gurele, R 077125, Romania}

\begin{abstract}

We address the problem of calculating the correlation functions of one-dimensional two-component
gases with strong repulsive contact interactions. The model considered in this paper describes
particles with fractional statistics and in appropriate limits reduces to the Gaudin-Yang model
or the spinor Bose gas. In the case of impenetrable particles  we derive a Fredholm determinant
representation for  the temperature-, time-, and space-dependent correlation functions which is
very easy to implement numerically and constitute the starting point for the analytical investigation
of the asymptotics.  Making use of this determinant representation and the solution of an associated
Riemann-Hilbert problem we derive the low-energy asymptotics of the correlators in the spin-incoherent
regime characterized by near ground-state charge degrees of freedom but a highly thermally disordered
spin sector. The asymptotics present features reminiscent of spin-charge separation with the spin part
exponentially decaying  in space separation and oscillating with a period proportional to the statistics
parameter while the charge part presents scaling with anomalous exponents which cannot be described by
any unitary conformal field theory. The momentum distribution and the Fourier transform of the dynamical
Green's function are asymmetrical for arbitrary statistics, a  direct consequence of the broken space-reversal
symmetry. Due to the exponential decay the momentum distribution $n(k)$ at zero temperature does not present
algebraic singularities  but the  tails obey the universal decay $\lim_{k\rightarrow\pm\infty}n(k)\sim C/k^4$
with the amplitude $C$  given by Tan's  contact. As a function of the statistics parameter the contact is a
monotonic function reaching its minimum for the fermionic system and the maximum for the bosonic system.

\end{abstract}

\maketitle

\onecolumngrid

\section{Introduction}

Many interacting one-dimensional (1D) systems are described at low energy by an effective theory known as
Tomonaga-Luttinger liquid (TLL) theory \cite{Hald1,Hald2,Giam}. This theory is expressed in terms of
many-body  collective excitations propagating with different velocities giving rise to an exotic behavior
known as  spin-charge separation. Tomonaga-Luttinger liquids are characterized by linear dependence on
temperature of the specific heat, algebraic decay of correlation functions including the single-particle
Green's function  and power-law vanishing of the tunneling  density of states as the chemical potential
is approached. The properties  of the TLL phase can be understood from bosonisation which maps the TLL
onto a theory of free massless compactified  bosonic fields with the compactified radii playing the role
of the phenomenological parameters that define the  effective theory \cite{FMS}. In the case of exactly
solvable many-body systems these phenomenological constants can  be extracted form the thermodynamic
properties of the system which can be computed using the thermodynamic Bethe ansatz.  The TLL description
is valid when both the charge and spin sectors of the system are at low energy. However, for strongly
interacting systems at low densities we can encounter the situation in which the temperature is much larger
than the single excitation spin energy but is much smaller than the single excitation for charge. This
energy window  $E_{spin}\ll k_BT\ll E_{charge}$ in which the spin energy is exponentially smaller than the
charge energy characterizes the spin-incoherent Tomonaga-Luttinger liquid (SITLL) regime \cite{BL,B1,CZ1,
CZ2,Matv,FB,CSZ,F} which has distinct properties and a higher degree of universality than the TLL. In this paper
we are going to investigate  the correlation functions of two-component anyonic gases with $\delta$-function
interaction in the spin-incoherent regime.

Integrable models play an important role in our understanding of strongly interacting 1D physical systems
\cite{KBI,EFGKK}.  In principle the knowledge of their wave functions and energy spectrum provide the starting
point in computing  the correlation functions from first principles without resorting to sometimes hard to
justify approximations. In practice, the complexity of the Bethe ansatz wave functions means that such
calculations are extremely difficult and new mathematical methods need to be devised. Initial attempts focused
on the case of systems equivalent with free fermions \cite{BL,B1,CZ1,CZ2,Len,VT1,VT2,Gang,JMMS,IIK2,IIK3,IIKV,
IIKS2,GIKP,GIK,PKA4,PKA5} but very recently results were obtained for the Lieb-Liniger \cite{KKMST2,KKMST1,
KKMT, KMS1,KMS2,Koz3,KT1,PK3,CC1,CCS,PC0} and the XXZ spin-chain \cite{KMBFM,ABS,CM0,CH0,BDGKS,TGK,SABGK,DGK,DGKS1,DGKS2,
GGKS,K4,K3}  models away from the free fermion point. Analytical derivations of the low-energy asymptotics of
correlation  functions are important because they can be compared with the predictions of TLL but they can also
provide insight  and identify systems not described by TLL theory. An important example is represented by the
Gaudin-Yang model  \cite{Y1,G1} which describes 1D nonrelativistic fermions with $\delta$-function repulsive
interaction (see \ref{ham}). At finite coupling strength $c$ and low energies the model is well described by
TLL theory \cite{Giam} and can be understood as the continuum limit of the Hubbard model \cite{EFGKK}. The case
of impenetrable particles,  $c\rightarrow\infty$, is particularly interesting due to the infinite spin degeneracy
of the ground state. For this reason the TLL description fails and the effect on the correlation functions on
taking this limit is highly  nontrivial  due to the noncommutativity of $c\rightarrow
\infty$ and $T\rightarrow 0$.  Taking the limit of vanishing temperature first, followed by  $c\rightarrow\infty$,
results are obtained which are  consistent with the impenetrable limit of the TLL or Conformal Field Theory (CFT)
description. The static single-particle  Green's function decays algebraically and the momentum distribution behaves
like $n(k)\sim n_{k_F}-\mbox{const}|k-k_F|^{1/8} \mbox{sgn}(k-k_F)$ near $k_F$ \cite{RA}. However, by taking the
impenetrable limit first and then $T\rightarrow 0$ the static Green's function is exponentially decaying in space
separation and the momentum distribution is no longer singular at $k_F$. The importance of this result derived by
Berkovich and Lowenstein \cite{BL,B1}  was overlooked for a long period of time. In 2004 Cheianov and Zvonarev
\cite{CZ1,CZ2} computed the low energy asymptotics of both static
and  dynamic correlators using the Deift-Zhou nonlinear steepest descent method \cite{DZ,Deift} and showed that they do not fit the
TLL predictions. The asymptotics show signs of spin-charge separation with the spin part exponentially decaying
and the charge part presenting scaling behavior with anomalous exponents inconsistent with any unitary conformal theory.
These features reveal that the correlators of the Gaudin-Yang model derived by taking the successive limits $c\rightarrow\infty$ and
$T\rightarrow 0$ belong to the spin-incoherent Tomonaga-Luttinger liquid regime \cite{CZ1,CZ2,Matv,FB,CSZ,F}.

In this paper we investigate the correlation functions of a 1D two-component  system with repulsive contact
interactions which can be understood as the anyonic generalization of the Gaudin-Yang model. The
literature on 1D anyonic models has been growing steadily in the last years \cite{AN,Girar,CM,LMP,IT1,IT2,BGK,BCM,Greit,Kundu2,CMT,
Belazz,BS,RFB,WWZ,MS,LH,DKPBJ,RCSM,HK,YP,HZC2,HC1,Hao1,ZGFSZ,AFFSV1,AFFSV2} spurred by several experimental proposals
of realizing such systems with  ultracold atoms in optical lattices using Raman-assisted tunneling \cite{KLMR,GS},
periodically driven lattices \cite{SSE} or multicolor lattice-depth modulation \cite{CGS1,GCS2}. Much of the attention
was focused on single component systems like the anyonic Lieb-Liniger model \cite{Kundu,BGO,BGH}  for which we know
the ground state characterization \cite{BGO,HZC1}, asymptotic behavior of the correlation functions for homogeneous
\cite{CM,PKA,SC,CS1,PKA4,PKA5} and trapped systems \cite{MPC,HS}, entanglement \cite{SFC,GHC}, and  nonequilibrium
\cite{delC,WRDK,PC} properties.  In the case of multi-component anyonic systems, such as the one considered
in this paper, there are fewer known results, especially from the analytical point of view \cite{YCLF,SPK,Zinn}.
The starting point of our analysis is the derivation of the  Fredholm determinant representation for the temperature-,
time-, and space-dependent single-particle Green's function in the strongly interacting regime using the summation of
form factors. This representation is obtained by taking the limit $c\rightarrow\infty$ first in the wavefunctions
which means that the correlators in the zero temperature limit will be in the spin-incoherent regime. This constitutes
the anyonic generalization of the fermionic and bosonic result obtained by Izergin and Pronko \cite{IP}. The infrared
asymptotics of the correlators at zero temperature and zero magnetic field are computed using the solutions of the associated
Riemann-Hilbert problems to the static and time generalization of the sine-kernel \cite{KKMST,K2}.  The spin part of
the asymptotics is exponentially decaying (the correlation length is independent on statistics) coupled with an
oscillatory component with frequency depending on the statistics parameter while the charge part presents scaling with
anomalous exponents. These features (except the statistics dependent oscillatory part which is a  defining feature
of 1D anyonic systems) show that the system is in the spin-incoherent regime. For the fermionic system not only  we
recover Cheianov and Zvonarev \cite{CZ1,CZ2} results but we are also able to compute the constants in front of the
asymptotics in the time-like regime. As a by-product of our analysis we also derive the zero temperature asymptotics for
the correlators  of single component anyons (Lieb-Liniger anyons) \cite{Kundu, BGO,PKA}. The momentum distribution is
not symmetric but unlike the single-component
counterpart it does not present a singularity at $(1-\kappa)k_F$ ($\kappa\in[0,1]$ is the statistics parameter) due to
the exponential decay of the static correlator. The high-momentum tails present the universal $\lim_{k\rightarrow
\pm\infty}n(k)\sim C/k^4$ behavior of models with contact interactions with the amplitude $C$ called  Tan's contact
(it should be pointed out that in the case of penetrable anyons this might not necessarily be true). Similar to the
momentum distribution the Fourier transform of the field-field correlator is not symmetric due to the broken
space-reversal symmetry.

The plan of the paper is as follows. In Sect.~\ref{s2} we introduce the model, its eigenstates, spectrum,  and Bethe
ansatz equations. The general form of the determinant representation for the correlators and some particular limits
(static, zero temperature, single-component) are presented in Sect.~\ref{s3}. The large distance asymptotic behavior
of the static correlators at zero temperature and zero magnetic field, momentum distribution and contact are investigated
in Sect.~\ref{s5}. The low energy asymptotics of the dynamic correlators in both space-like and time-like regions are
calculated in  Sect.~\ref{s6} and similar results for the single component system are reported in Sect.~\ref{s5b}.
The derivation of the determinant representation is succinctly presented in
Sects.~ \ref{s6}, \ref{s7}, \ref{s8}. In Sect.~\ref{s6} we compute the form factors, in Sect.~\ref{s7} we summate them
and the thermodynamic limit is taken in Sect.~\ref{s8}. We conclude in Sect.~\ref{s9}. Some minimal information on Fredholm
determinants, their numerical implementation, the asymptotic solution of certain Riemann-Hilbert problems and the rewriting of relevant
functions in the thermodynamic limit are presented in several appendices.

\section{Model and eigenstates}\label{s2}

We consider a one-dimensional two-component system of anyonic particles with  spin-independent
repulsive contact interactions. For a finite system of length $L$ the Hamiltonian in second
quantization is (we use units of $\hbar=2m=k_B=1$)
\be\label{ham}
\mathcal{H}=\int_{0}^L dx\, \left[ (\6_x \bm{\psi^\dagger} \6_x \bm{\psi})+c:(\bm{\psi^\dagger}\bm{\psi}):
-h (\bm{\psi^\dagger}\bm{\psi})+B (\bm{\psi^\dagger}\sigma_z\bm{\psi})\right]\, ,
\ee
with $c>0$ the coupling strength, $h$ the chemical potential, $B$ the magnetic field and
$:\,\, :$ denote normal ordering. In (\ref{ham}) $\bm{\psi^\dagger}(x)=\left(\psi_1^\dagger(x)
\, ,\psi_2^\dagger(x)\right)\, ,$ $\bm{\psi}(x)=\left(\psi_1(x)\, ,\psi_2(x)\right)^T$
with $\sigma_z$ the third Pauli matrix and $\psi_\alpha(x)$ are anyonic field operators satisfying
anyonic commutation relations
\begin{subequations}\label{commr}
\begin{align}
\psi_\alpha(x)\psi_\beta^\dagger(y)&=-e^{-i \pi\kappa\, \scriptsize{\mbox{sgn}}(x-y)}\psi_\beta^\dagger(y)\psi_\alpha(x)+\delta_{\alpha,\beta}\delta(x-y)\, ,\\
\psi_\alpha(x)\psi_\beta(y)&=-e^{i \pi\kappa\, \scriptsize{\mbox{sgn}}(x-y)}\psi_\beta(y)\psi_\alpha(x)\, ,\\
\psi_\alpha^\dagger(x)\psi_\beta^\dagger(y)&=-e^{i \pi\kappa\, \scriptsize{\mbox{sgn}}(x-y)}\psi_\beta^\dagger(y)\psi_\alpha^\dagger(x)\, ,
\end{align}
\end{subequations}
where $\kappa\in[0,1]$ is the statistics parameter and $\mbox{sgn}(x)=x/|x|$ with $\mbox{sgn}
(0)=0$. Varying the statistics parameter in the interval $[0,1]$ the anyonic commutation
relations interpolate continuously between the ones for fermions $\kappa=0$  and bosons
$\kappa=1$. At $\kappa=0$ the Hamiltonian (\ref{ham}) becomes the integrable
Gaudin-Yang model \cite{Y1,G1} and at $\kappa=1$ it describes the spinor Bose gas which is
also integrable \cite{S1,Kul1,KR,LGYU,Slav1}.  For an arbitrary value of $\kappa$ and
coupling strength $c$ the solution of (\ref{ham}) can in principle be obtained employing the
quantum inverse scattering method with anyonic grading \cite{BFGLZ}. In this paper we are
interested in the limit of infinite repulsion, $c\rightarrow\infty$, which is simpler and
for which the eigenfunctions can be determined in a more easier fashion. An observation is
in order. While we have chosen $\kappa\in[0,1]$ an equally valid choice would have been
$\kappa\in[-1,0]$ with the bosonic system described by $\kappa=-1$.

The time- and space-dependent Green's functions (field-field correlators) at finite
temperature in the presence of a magnetic field are defined as
\begin{align}\label{defcorr}
\Gm{\beta}(x,t\,|\,\kappa,T,B,h)&\equiv\langle\psi^\dagger_\beta(x,t)\psi_\beta(0,0)\rangle_{\kappa,T,B,\,h}=
\frac{\mbox{Tr}\left[e^{-\mathcal{H}/T}\psi^\dagger_\beta(x,t)\psi_\beta(0,0)\right]}{\mbox{Tr}\left[e^{-\mathcal{H}/T}\right]}\, ,\ \ \beta \in\{1,2\}\, , \\
\Gp{\beta}(x,t\,|\,\kappa, T,B,h)&\equiv\langle\psi_\beta(x,t)\psi_\beta^\dagger(0,0)\rangle_{\kappa,T,B,\,h}=
\frac{\mbox{Tr}\left[e^{-\mathcal{H}/T}\psi_\beta(x,t)\psi^\dagger_\beta(0,0)\right]}{\mbox{Tr}\left[e^{-\mathcal{H}/T}\right]}\, ,\ \ \beta \in\{1,2\}\, ,
\end{align}
where $\psi_\beta^\dagger(x,t)=e^{i t \mathcal{H}} \psi_\beta^\dagger(x) e^{-i t \mathcal{H}}
\, ,$ $\psi_\beta(x,t)=e^{i t \mathcal{H}} \psi_\beta(x) e^{-i t \mathcal{H}}$ and the trace is
taken over the Fock space. The correlators of the second type of particles are related to the
ones of the first type of particles by changing the sign of the magnetic field
\be\label{symm}
\mathcal{G}_{1}^{\mbox{\scriptsize{($\pm$)}}}(x,t\,|\,\kappa,T,B,h)=\mathcal{G}_{2}^{\mbox{\scriptsize{($\pm$)}}}(x,t\,|\,\kappa,T,-B,h)\, .
\ee
so in principle it is sufficient to investigate only  $\mathcal{G}_{1}^{(\pm)}(x,t)$ (here,
and in the following, we dropped the dependence on certain parameters when there is no risk
of confusion). In order to study the correlators in the thermodynamic limit we will consider
first the finite size system and then will take the limit $L\rightarrow\infty$ keeping the
densities of both type of particles fixed.

\subsection{Wavefunctions and eigenvectors}

We say that a state of the system is in the $(N,M)$-sector if the total number of particles is
$N$ of which $M$ are of type 2 ($\alpha=2$). The Hamiltonian (\ref{ham}) conserves the number
of each type of particles separately and an eigenstate in the $(N,M)$-sector can be written as
$(\bm {z}=\{z_1,\cdots,z_N\}\, ,\ \bm {\alpha}=\{\alpha_1,\cdots,\alpha_N\}$)
\be\label{lab1}
|\Psi_{N,M}(\bm{k},\bm{\lambda})\rangle=\int_0^L d z_1 \cdots   d z_N \sum_{\alpha_1,\cdots,\alpha_N=\{1,2\}}^{[N,M]}\chi_{N,M}^{\bm{\alpha}}
(\bm{z}|\bm{k},\bm{\lambda})\, \psi_{\alpha_N}^\dagger(z_N)\cdots\psi_{\alpha_1}^\dagger(z_1)|0\rangle\, ,
\ee
with $\bm{k}=\{k_j\}_{j=1}^N$ and $\bm{\lambda}=\{\lambda_j\}_{j=1}^M$ two sets of unequal
parameters (note the ordering of the creation operators which helps the subsequent calculations
\cite{PKA}). In the upper limit of the sum appearing in (\ref{lab1}) $[N,M]$ means that $N-M$ of
the $\alpha$'s are equal to 1 and $M$ are equal to 2. $|0\rangle$ is the anyonic Fock vacuum which
satisfies $\langle0|0\rangle=1\, ,\ \psi_\alpha(z)|0\rangle=0$ and $0=\langle 0|\psi^\dagger_
\alpha(z)\, .$

In order to determine the wavefunctions we are going to use the same method as the one employed by
Izergin and Pronko \cite{IP} in the case of two-component fermions and bosons. Namely, first we are
going to take the limit $c\rightarrow\infty$ in the wavefunctions of the infinite line and then impose
periodic boundary conditions
\footnote
{
We should point out that imposing periodic boundary conditions in the case of 1D anyonic systems is
not completely trivial as it was first pointed out by Averin and Nesteroff \cite{AN} (see also
Appendix A of \cite{PKA}).
}
\be
\chi_{N,M}^{\bm{\alpha}}(0,z_2,\cdots,z_N|\bm{k},\bm{\lambda})=\chi_{N,M}^{\bm{\alpha}}(L,z_2,\cdots,z_N|\bm{k},\bm{\lambda})\, .
\ee
The first step produces the wavefunctions
\be\label{wavef}
\chi_{N,M}^{\alpha_1\cdots\alpha_N}(\bm{z}|\bm{k},\bm{\lambda})=\frac{1}{N!}
\left(\sum_{P\in S_N}\xi_{N,M}^{\alpha_{P(1)}\cdots\alpha_{P(N)}}(\bm{\lambda})e^{i\frac{\pi\kappa}{2}\sum_{1\le a<b\le N}\scriptsize{\mbox{sgn}}(z_a-z_b)}
\theta(z_{P(1)}<\cdots<z_{P(N)})\right)\mbox{det}_N e^{i k_a z_b},
\ee
where $S_N$ is the group of permutations of $N$ elements, $\theta(z_{P(1)}<\cdots<z_{P(N)})$ is a function which is equal to one when
$z_{P(1)}<\cdots<z_{P(N)}$ and zero otherwise, and $\mbox{det}_N e^{i k_a z_b}$ is the determinant of the $N\times N$ matrix with elements
$e^{i k_a z_b}\, .$
The $\xi_{N,M}^{\alpha_1 \cdots\alpha_N}(\bm{\lambda})$s are components of a $2^N$-dimensional vector  $\xi_{N,M}(\bm{\lambda})$ and
in principle the components can be chosen at will as long as they produce a complete set of eigenstates $|\Psi_{N,M}(\bm{k},\bm{\lambda})\rangle$
with the required symmetry. Following \cite{IP} we are going to choose $\xi_{N,M}(\bm{\lambda})$  as the eigenvectors of the XX0 spin chain with
periodic boundary conditions and $N$ spins of which $M$ are down. Explicitly, the components of $\xi_{N,M}(\bm{\lambda})$ are \cite{CIKT}
\be\label{dxi}
\xi_{N,M}^{\alpha_1\cdots\alpha_N}(\bm{\lambda})=\left(\prod_{1\le j<l\le N}\mbox{sgn}(n_l-n_j)\right)\mbox{det}_M e^{i \lambda_a n_b}\, ,
\ee
where $n_j$ gives the position of the $j$-th particle of type 2 in  $\{\alpha_1,\cdots,\alpha_n\}$ and the rapidities satisfy
\be\label{BAE2}
e^{i\lambda_l N}=(-1)^{M+1}\, ,\ \ \l=1,\cdots,M\, .
\ee
The wavefunctions (\ref{wavef}) present the required anyonic symmetry when interchanging two particles of the same type
($\xi_{N,M}^{\bm{\alpha}}(\bm{\lambda})$ are symmetric in $n$'s)
\be
\chi_{N,M}^{\bm{\alpha}}(\cdots, z_i,z_{i+1},\cdots|\bm{k},\bm{\lambda})=-e^{i\pi\kappa\, \scriptsize{\mbox{sgn}}(z_i-z_{i+1})}
\chi_{N,M}^{\bm{\alpha}}(\cdots, z_{i+1},z_{i},\cdots|\bm{k},\bm{\lambda})\, .
\ee
Imposing now periodic boundary conditions on $\chi_{N,M}^{\bm{\alpha}}(\bm{z}|\bm{k},\bm{\lambda})$ we find
\be\label{lab2}
\xi_{N,M}^{\alpha_1\cdots\alpha_N}(\bm{\lambda})=\xi_{N,M}^{\alpha_2\cdots\alpha_N\alpha_1}(\bm{\lambda})e^{i\pi\kappa(N-1)}e^{i k_j L}\, ,\ \ j=1,\cdots,N\, ,
\ee
which can be interpreted as the eigenvalue problem for the cyclic shift operator $C_N$ acting in a $2^N$-dimensional
vector space. Eq.~(\ref{lab2}) yields
\be\label{BAE1}
e^{ik_j L}=e^{-i\pi\kappa(N-1)}e^{i\sum_{b=1}^M \lambda_b}\, ,\ \ j=1,\cdots,N\, .
\ee
Eqs.~(\ref{BAE2}) and (\ref{BAE1}) represent the Bethe ansatz equations (BAEs) of the system. The states (\ref{lab1}) with
wavefunctions $\chi_{N,M}^{\bm{\alpha}}(\bm{z}|\bm{k},\bm{\lambda})$ defined in (\ref{wavef}) and $\xi_{N,M}^{\bm{\alpha}}(\bm{\lambda})$
given by (\ref{dxi}) represent a complete set of eigenstates of the Hamiltonian (\ref{ham}) provided that $\bm{k}$ and $\bm{\lambda}$
satisfy the BAEs. The normalization of the eigenstates is
\be\label{normalization}
\langle \Psi_{N,M}(\bm{k},\bm{\lambda})|\Psi_{N',M'}(\bm{k'},\bm{\lambda'})\rangle=
N^M L^N\delta_{N,N'}\delta_{M,M'}\delta_{\bm{k},\bm{k'}}\delta_{\bm{\lambda},\bm{\lambda'}}\, .
\ee
The allowed values for the quasimomenta $[k_a]_j\in(-\infty,+\infty)$ and $[\lambda_b]_l\in(-\pi,\pi]$ are
\begin{subequations}\label{allowed}
\begin{align}
[k_a]_j=&\frac{2\pi}{L}(j+\delta)+\frac{1}{L}\sum_{b=1}^M\lambda_b\, ,\ \ \ j\in\mathbb{Z}\, ,\ \ \delta=\{[-i\pi\kappa(N-1)]\}\, ,\ \  \ \ (a=1,\cdots,N)\, ,\\
[\lambda_b]_l=&\frac{2\pi}{N}\left(-\frac{N}{2}-\frac{1+(-1)^{N-M}}{4}+l\right)\, ,\ \ \ l=1,\cdots,N\, , \ \ (b=1,\cdots,M)\, ,
\end{align}
\end{subequations}
where we have introduced the notation $\{[x]\}=\gamma$ if $x=2\pi \times\mbox{integer}+2\pi \gamma$ with $\gamma\in(-1/2,1/2]\, .$
The eigenvalues of the Hamiltonian (\ref{ham}) $\mathcal{H}|\Psi_{N,M}(\bm{k},\bm{\lambda})\rangle=E_{N,M}(\bm{k}) |\Psi_{N,M}(\bm{k},\bm{\lambda})\rangle$ are
\be\label{energy}
E_{N,M}(\bm{k})=\sum_{j=1}^N(k_j^2-h+B)-2MB\, .
\ee

\section{Determinant representation for the correlators}\label{s3}

Knowledge of the eigenstates and the exact form of the wavefunctions opens the way for the computation of the correlators
using a method first introduced by Korepin and Slavnov in the case of the impenetrable Bose gas \cite{KS}.
We start with a finite system and insert a resolution of identity between the field operators  appearing on the right hand side of
(\ref{defcorr}) obtaining a sum over form factors.
Fortunately,  the form factors can be computed relatively easily  and  the summation over them can be performed using a procedure
which can be called the ``insertion of the summation into the determinant "\cite{KS,IP}. Taking the thermodynamic limit in (\ref{defcorr})
produces Fredholm determinants which can be easily computed numerically and are particularly suited to asymptotic analysis.
The entire derivation is presented  in Sections \ref{s6}, \ref{s7} and \ref{s8}. In this section we present the final results.

The temperature-, time-, and space-dependent single-particle Green's functions (\ref{defcorr}) admit the following Fredholm determinant representations
(for the definition of the Fredholm determinant see Appendix \ref{a1}):
\begin{align}
\Gm{1}(x,t\,|\,\kappa,T,B,h)&=e^{- i t(h-B)}\int_{-\pi}^\pi \frac{d\eta}{2\pi}\,
F(\gamma,\eta)\left[\mbox{\textsf{det}}\left(1+\gamma\, \hat{\textsf{V}}^{\mbox{\scriptsize{(T,B,--)}}}(\eta)+
\hat{\textsf{R}}^{\mbox{\scriptsize{(T,B,--)}}}\right)
- \mbox{\textsf{det}}\left(1+\gamma\, \hat{\textsf{V}}^{\mbox{\scriptsize{(T,B,--)}}}(\eta) \right)\right]\, ,\label{detm}\\
\Gp{1}(x,t\,|\,\kappa,T,B,h)&=e^{ i t(h-B)}\int_{-\pi}^\pi \frac{d\eta}{2\pi}\,
F(\gamma,\eta)\left[\mbox{\textsf{det}}\left(1+\gamma\, \hat{\textsf{V}}^{\mbox{\scriptsize{(T,B,+)}}}(\eta)-
\gamma\,\hat{\textsf{R}}^{\mbox{\scriptsize{(T,B,+)}}}(\eta)\right)\right.\nonumber\\
&\qquad\qquad\qquad\qquad\qquad\qquad\qquad\qquad\qquad\qquad\qquad
\left.+(G(x,t)-1)\,  \mbox{\textsf{det}}\left(1+\gamma\, \hat{\textsf{V}}^{\mbox{\scriptsize{(T,B,+)}}}(\eta) \right)\right]\, ,\label{detp}
\end{align}
where
$\gamma=1+e^{2 B/T}\, , $ $F(\gamma,\eta)=1+\sum_{p=1}^\infty \gamma^{-p} (e^{i\eta p}+e^{-i \eta p})\, $ and $G(x,t)=\inti  e^{-i t k^2+i kx}\, dk/2\pi .$
The action of the integral operators $\hat{\textsf{V}}^{\mbox{\scriptsize{(T,B,$\pm $)}}}$ and
$\hat{\textsf{R}}^{\mbox{\scriptsize{(T,B,$\pm$)}}}$   on an arbitrary function $f(k)$ is given by
\be
\left(\hat{\textsf{V}}^{\mbox{\scriptsize{(T,B,$\pm $)}}}f \right)(k)=\inti \textsf{V}^{\mbox{\scriptsize{(T,B,$\pm $)}}}(\eta,\kappa\, |\, k,k')f(k')\, dk'\, \ \ \
\left(\hat{\textsf{R}}^{\mbox{\scriptsize{(T,B,$\pm $)}}}f \right)(k)=\inti \textsf{R}^{\mbox{\scriptsize{(T,B,$\pm $)}}}(\eta,\kappa\, |\, k,k')f(k')\, dk'\, ,
\ee
with kernels
\begin{align}
\textsf{V}^{\mbox{\scriptsize{(T,B,$\pm $)}}}(\eta,\kappa\, |\, k,k')&=4 \sin^2 \pi \nu^{(\pm)}\, \sqrt {\vartheta(k)}\,
\frac{E^{(\pm)}(k)\,e(k')-E^{(\pm)}( k')\,e(k)}{2\pi i (k-k')}\, \sqrt {\vartheta(k')}\, ,\ \ \ \nu^{(\pm)}=\pm\left(\frac{\eta}{2\pi}-\frac{\kappa}{2}\right)\, ,\\
\textsf{R}^{\mbox{\scriptsize{(T,B,+)}}}(\eta,\kappa\, |\, k,k')&=-2 \sin^2 \pi \nu^{(+)}\, \sqrt {\vartheta(k)}\,
\frac{E^{(+)}(k)E^{(+)}(k')}{\pi} \, \sqrt {\vartheta(k')}\, ,\\
\textsf{R}^{\mbox{\scriptsize{(T,B,--)}}}(\eta,\kappa\, |\, k,k')\, &= \sqrt {\vartheta(k)}\,\,
\frac{e(k) e(k')}{2\pi} \, \sqrt {\vartheta(k')}\, , \ \ \ \vartheta(k)=\frac{e^{-B/T}}{2\cosh(B/T) +e^{(k^2-h)/T}}\, .
\end{align}
Here $\vartheta(k)$ is the Fermi function and
\begin{subequations}
\begin{align}
e(x,t\, |\,k)&=e^{\frac{i t k^2}{2}- \frac{i k x}{2}}\, ,\label{defe}\\
E^{(\pm)}(x,t\, |\, k)&= ie(k)\left\{\mbox{p.v.} \inti \frac{dq}{2\pi} \frac{e^{-2}(x,t\,|\,q)}{q-k}+\frac{1}{2}\cot\pi\nu^{(\pm)}e^{-2}(x,t\, |\,k)\right\}\, ,\label{defE}
\end{align}
\end{subequations}
with p.v. denoting the principal value integral.
The representations for the correlators $\mathcal{G}^{(\pm)}_2(x,t\,|\,\kappa,T,B,h)$ are obtained from (\ref{detm}) and (\ref{detp}) using the symmetry (\ref{symm}).
Other useful symmetries are (the bar stands for complex conjugation)
\be\label{gensymm}
\mathcal{G}^{(\pm)}_\alpha(x,t\,|\,\kappa)=\overline{\mathcal{G}}^{(\pm)}_\alpha(-x,-t\,|\,\kappa)\, ,\ \
\mathcal{G}^{(\pm)}_\alpha(-x,t\,|\,\kappa)=\mathcal{G}^{(\pm)}_\alpha(x,t\,|-\kappa)\,  \mbox{and}\ \
\mathcal{G}^{(\pm)}_\alpha(x,-t\,|\,\kappa)=\overline{\mathcal{G}}^{(\pm)}_\alpha(x,t\,|-\kappa)\, .
\ee
Our results are equivalent with the determinant representations derived by Izergin and Pronko \cite{IP} for  impenetrable
two-component fermions ($\kappa=0$) and two-component bosons ($\kappa=1$) but we should point out a subtle difference. If
we would rewrite the results from \cite{IP} in our notation their $E^{(-)}(k)$ function for fermions has a plus sign instead
of a minus sign in front of the  $\cot(\eta/2)$ term (see Eq.~(\ref{defE})  for $\kappa=0$). However this minus sign  is
irrelevant because in the expansion of the determinant it will give contributions of the type $(-\cot(\eta/2))^p$ integrated
over a symmetric interval and therefore only even powers of $p$ will give a contribution. In the rest of this section we are
going to present some particular cases of the general formulae (\ref{detm}) and (\ref{detp}).

\subsection{Static correlators}

The determinant representations for the correlators simplify considerably in the static limit. For $t=0$  the principal value integral
appearing in Eq.~(\ref{defE}) can be computed analytically by closing the contour in the upper half-plane for $x>0$ and in the lower half plane
for $x<0$ with the result
$
\mbox{p.v.} \inti  e^{i q x}/2\pi (q-k)\, dq= i\, \mbox{sgn} (x)\, e^{i k x}/2\,  .
$
The relevant functions $e(k)$ and $E^{(\pm)}(k)$ reduce to
$e(x,0\, |\,k)=e^{- \frac{i k x}{2}}\, ,$
$ E^{(\pm)}(x,0\, |\, k)= e^{- \frac{i k x}{2}}\left[i \cot\pi\nu^{(\pm)}-\mbox{sgn}(x)\right]/2 \, .$
Introducing the integral operators  $\hat{\textsf{v}}^{\mbox{\scriptsize{(T,B,$\pm $)}}}$ and
$\hat{\textsf{r}}^{\mbox{\scriptsize{(T,B,$\pm$)}}}$  acting on the entire real axis with kernels
\begin{align}
\textsf{v}^{\mbox{\scriptsize{(T,$\pm $)}}}(k,k')&= -2\, \sqrt {\vartheta(k)}\,\,
 \frac{\sin[(k-k') |x|/2)]}{\pi  (k-k')}\, \sqrt {\vartheta(k')}\, ,\label{defv}\\
\textsf{r}^{\mbox{\scriptsize{(T,$\pm$)}}}( k,k')&=\frac{1}{2\pi} \sqrt {\vartheta(k)}\,  e^{\pm i(k+k') x/2}
 \, \sqrt {\vartheta(k')}\, ,\label{defr}
\end{align}
then $
\textsf{V}^{\mbox{\scriptsize{(T,B,$\pm $)}}}(\eta,\kappa\, |\, k,k') \overset{t=0}{\rightarrow}\frac{1}{2}\left(1-e^{\mp i\, \scriptsize{\mbox{sgn}} (x)\, \pi \kappa} e^{\pm i \eta}\right)\, \textsf{v}^{\mbox{\scriptsize{(T,$\pm $)}}}(k,k')\, ,$
$
\textsf{R}^{\mbox{\scriptsize{(T,B,+)}}}(\eta,\kappa\, |\, k,k') \overset{t=0}{\rightarrow} e^{i\, \scriptsize{\mbox{sgn}} (x)\, (\eta-\pi \kappa)}\textsf{r}^{\mbox{\scriptsize{(T,+)}}}( k,k')\,
$
and
$
\textsf{R}^{\mbox{\scriptsize{(T,B,--)}}}(\eta,\kappa\, |\, k,k') \overset{t=0}{\rightarrow} \textsf{r}^{\mbox{\scriptsize{(T,--)}}}( k,k')\, .
$
Because the dependence on $\eta$ is now very simple we can integrate. For example, expanding  the first determinant appearing in the right hand side of
(\ref{detm}) for $t=0$ and $x>0$ we have $\mbox{\textsf{det}}\left(1+\gamma\, \hat{\textsf{V}}^{\mbox{\scriptsize{(T,B,--)}}}(\eta)+
\hat{\textsf{R}}^{\mbox{\scriptsize{(T,B,--)}}}\right) \overset{x=0}{\rightarrow} \sum_{n=0}^\infty (\frac{\gamma}{2})^n \left(1-e^{i \pi \kappa}e^{-i \eta}\right)^n
A_n(k,k')$ where $A_n(k,k')$ are coefficients that do not depend on $\eta$. This term will give the contribution
\be
 \int_{-\pi}^\pi \frac{d\eta}{2\pi}\left(1+\sum_{p=1}^\infty\gamma^{-p}(e^{ip \eta}+e^{-ip\eta})\right) \sum_{n=0}^\infty \left(\frac{\gamma}{2}\right)^n \left(1-e^{i \pi \kappa}e^{-i \eta}\right)^n
A_n(k,k')
=\sum_{n=0}^\infty \left(\frac{\gamma}{2}-\frac{e^{i\pi\kappa}}{2}\right)^n A_n(k,k')\, ,
\ee
where we have use the binomial formula and integrated term by term. Therefore the determinant representation for the static $\mathcal{G}^{\mbox{\scriptsize{(--)}}}_1(x,0)$ correlator
is given by
\be\label{corrsm}
\Gm{1}(x,0)=
\mbox{\textsf{det}}\left(1+\tilde{\gamma}\, \hat{\textsf{v}}^{\mbox{\scriptsize{(T,--)}}}+
\hat{\textsf{r}}^{\mbox{\scriptsize{(T,--)}}}\right)
- \mbox{\textsf{det}}\left(1+\tilde{\gamma}\,\hat{\textsf{v}}^{\mbox{\scriptsize{(T,--)}}} \right)\, ,\ \ \
\tilde{\gamma}=\left(\gamma-e^{i \, \scriptsize{\mbox{sgn}} (x)\,\pi\kappa}\right)/2\, ,
\ee
with $\textsf{v}^{\mbox{\scriptsize{(T,--)}}}(k,k')$ and $\hat{\textsf{r}}^{\mbox{\scriptsize{(T,--)}}}(k,k')$ defined in (\ref{defv}) and (\ref{defr}). In a similar
fashion it can be shown that ($\overline{\tilde{\gamma}}$ is the complex conjugate of $\tilde{\gamma}$)
\begin{align}\label{corrsp}
\Gp{1}(x,0)&=
\mbox{\textsf{det}}\left(1+\overline{\tilde{\gamma}}\,  \hat{\textsf{v}}^{\mbox{\scriptsize{(T,+)}}}
- e^{-i \, \scriptsize{\mbox{sgn}} (x)\,\pi\kappa}\, \hat{\textsf{r}}^{\mbox{\scriptsize{(T,+)}}}\right)
+(\delta(x)-1) \mbox{\textsf{det}}\left(1+\overline{\tilde{\gamma}}\, \hat{\textsf{v}}^{\mbox{\scriptsize{(T,+)}}} \right)\, .
\end{align}
In contrast with the fermionic ($\kappa=0$) and bosonic case ($\kappa=1$) the static correlators are complex
and they are different depending on the sign of  $x$. It is easy to see from the representations  (\ref{corrsm}) and (\ref{corrsp})
that we have
\be\label{staticsymm}
\mathcal{G}_{1}^{\mbox{\scriptsize{($\pm$)}}}(x,0)=\overline{\mathcal{G}}_{1}^{\mbox{\scriptsize{($\pm$)}}}(-x,0)\, .
\ee
As we will see below this results in a nonsymmetric momentum distribution. The $\mathcal{G}_{2}^{\mbox{\scriptsize{($\pm$)}}}(x,0)$
correlators are obtained using (\ref{symm}). The two types of correlators also satisfy $\Gp{1}(x,0)=\delta(x)-e^{-i\pi\kappa}\Gm{1}(-x,0)$.

\subsection{Correlators at $T=0$ and $B=0$}

At zero temperature the properties of the system are heavily influenced by the value of the magnetic field. This is due to
the fact that our model presents a quantum phase transition at $B=0$ with a ground state that is completely  polarized (only
particles of one type) for $|B|>0$ and a balanced one at $B=0$. The balanced case, including the asymptotic behavior of the
correlators and the momentum distribution, will be studied in detail in Sect. \ref{s4} and \ref{s5}. In the zero temperature
limit $\lim_{T\rightarrow 0} \gamma\vartheta(k)=\theta(k_F^2-k)$ with $\theta(k)$ the Heaviside function and $k_F=(h+|B|)^{1/2}.$
For the balanced system at zero temperature $\gamma=2$, $k_F=h^{1/2}$ and the correlators of both type of particles are equal  $\mathcal{G}_{1}^{\mbox{\scriptsize{($\pm$)}}}(x,t)=\mathcal{G}_{2}^{\mbox{\scriptsize{($\pm$)}}}(x,t).$
Using the identity 1.447 (3) of \cite{GR}, $1+2\sum_{p=1}^\infty a^p \cos p x=(1-a^2)/(1-2 a\cos x +a^2)\, , |a|<1$, we find
$F(\gamma=2,\eta)= 3/(5-4\cos \eta)$ and from (\ref{detm}) and (\ref{detp}) we obtain
\begin{subequations}\label{corr00}
\begin{align}
\Gm{1}(x,t)&=e^{- i th}\int_{-\pi}^\pi \frac{d\eta}{2\pi} \frac{3}{5-4\cos \eta}\,
\left[\mbox{\textsf{det}}\left(1+ \hat{\textsf{V}}^{\mbox{\scriptsize{(0,0,--)}}}(\eta)+
\frac{1}{2}\, \hat{\textsf{R}}^{\mbox{\scriptsize{(0,0,--)}}}\right)
- \mbox{\textsf{det}}\left(1+ \hat{\textsf{V}}^{\mbox{\scriptsize{(0,0,--)}}}(\eta) \right)\right]\, ,\label{detm0}\\
\Gp{1}(x,t)&=e^{+ i th}\int_{-\pi}^\pi \frac{d\eta}{2\pi} \frac{3}{5-4\cos \eta}
\left[\mbox{\textsf{det}}\left(1+ \hat{\textsf{V}}^{\mbox{\scriptsize{(0,0,+)}}}(\eta)-
\hat{\textsf{R}}^{\mbox{\scriptsize{(0,0,+)}}}(\eta)\right)\right.\nonumber\\
&\qquad\qquad\qquad\qquad\qquad\qquad\qquad\qquad\qquad\qquad\qquad
\left.+(G(x,t)-1)\,  \mbox{\textsf{det}}\left(1+ \hat{\textsf{V}}^{\mbox{\scriptsize{(0,0,+)}}}(\eta) \right)\right]\, ,\label{detp0}
\end{align}
\end{subequations}
where the integral operators  $\hat{\textsf{V}}^{\mbox{\scriptsize{(0,0,$\pm $)}}}$ and $\hat{\textsf{R}}^{\mbox{\scriptsize{(0,0,$\pm$)}}}$
act on $[-k_F,k_F]$  and have the kernels
\begin{subequations}\label{kernels00}
\begin{align}
\textsf{V}^{\mbox{\scriptsize{(0,0,$\pm $)}}}(\eta,\kappa\, |\, k,k')&=4 \sin^2 \pi \nu^{(\pm)}\,
\frac{E^{(\pm)}(k)\,e(k')-E^{(\pm)}( k')\,e(k)}{2\pi i (k-k')}\, ,\ \ \ \nu^{(\pm)}=\pm\left(\frac{\eta}{2\pi}-\frac{\kappa}{2}\right)\, ,\label{v0}\\
\textsf{R}^{\mbox{\scriptsize{(0,0,+)}}}(\eta,\kappa\, |\, k,k')&=-2 \sin^2 \pi \nu^{(+)}\,
\frac{E^{(+)}(k)E^{(+)} (k')}{\pi}\,  ,\ \ \
\textsf{R}^{\mbox{\scriptsize{(0,0,--)}}}(\eta,\kappa\, |\, k,k')\, =
\frac{e(k) e(k')}{2 \pi} \, .\label{rp0}
\end{align}
\end{subequations}
The functions $e(k)$ and $E^{(\pm)}( k)$ are defined in (\ref{defe}) and (\ref{defE}).

\subsection{Correlators at $T=0$ and $B\ne 0$}

We consider the case of negative magnetic field, $B<0$, keeping in mind that the correlators at positive magnetic
field can be determined using (\ref{symm}). The ground state is formed by particles of type 1,
$\gamma=1$ and $F(\gamma=1,\eta)=2\pi \delta(\eta)$. Therefore, we are effectively dealing with a single component system
(see Sect. \ref{scl}). The integration is now trivial and we find for the correlators
of type 1 particles
\begin{align}
\Gm{1}(x,t)&=e^{- i t(h-B)}
\left[\mbox{\textsf{det}}\left(1+ \hat{\textsf{V}}^{\mbox{\scriptsize{(0,B,--)}}}(0)+
\, \hat{\textsf{R}}^{\mbox{\scriptsize{(0,B,--)}}}\right)
- \mbox{\textsf{det}}\left(1+ \hat{\textsf{V}}^{\mbox{\scriptsize{(0,B,--)}}}(0) \right)\right]\, ,\\
\Gp{1}(x,t)&=e^{+ i t(h-B)}
\left[\mbox{\textsf{det}}\left(1+ \hat{\textsf{V}}^{\mbox{\scriptsize{(0,B,+)}}}(0)-
\hat{\textsf{R}}^{\mbox{\scriptsize{(0,B,+)}}}(0)\right)
+(G(x,t)-1)\,  \mbox{\textsf{det}}\left(1+ \hat{\textsf{V}}^{\mbox{\scriptsize{(0,B,+)}}}(0) \right)\right]\, ,
\end{align}
with the integral operators  $\hat{\textsf{V}}^{\mbox{\scriptsize{(0,B,$\pm $)}}}$ and $\hat{\textsf{R}}^{\mbox{\scriptsize{(0,B,$\pm$)}}}$
acting on $[-k_F,k_F]$ with $k_F=(h+|B|)^{1/2}$. The kernels are
\begin{align}
\textsf{V}^{\mbox{\scriptsize{(0,B,$\pm $)}}}(0,\kappa\, |\, k,k')&=4 \sin^2\left( \frac{\pi \kappa}{2}\right)\,
\frac{E^{(\pm)}(k)\,e(k')-E^{(\pm)}( k')\,e(k)}{2\pi i (k-k')}\, ,\ \ \ \nu^{(\pm)}=\mp\frac{\kappa}{2}\, ,\\
\textsf{R}^{\mbox{\scriptsize{(0,B,+)}}}(0,\kappa\, |\, k,k')&=-2 \sin^2\left( \frac{\pi \kappa}{2}\right)\,
\frac{E^{(+)}(k)E^{(+)}(k')}{\pi}\, ,\ \ \ \
\textsf{R}^{\mbox{\scriptsize{(0,B,--)}}}(0,\kappa\, |\, k,k')\, =
\frac{e(k) e(k')}{2 \pi} \, .
\end{align}
Because the ground state is comprised by only particles of type 1 the $\Gm{2}(x,t)$ correlator is zero. However, for the
other correlator we have $\Gp{2}(x,t\,|\,\kappa,0,B,h)=\Gp{1}(x,t\,|\,\kappa,0,-B,h)$ . For $B>0$ we have $\gamma=\infty$ and $F(\gamma,\eta)=1$,
therefore
\begin{align*}
\Gp{2}(x,t)&=e^{ i t(h+B)}\int_{-\pi}^\pi \frac{d\eta}{2\pi}
\left[\mbox{\textsf{det}}\left(1+ \hat{\textsf{V}}^{\mbox{\scriptsize{(0,0,+)}}}(\eta)-
\hat{\textsf{R}}^{\mbox{\scriptsize{(0,0,+)}}}(\eta)\right)+(G(x,t)-1)\,  \mbox{\textsf{det}}\left(1+ \hat{\textsf{V}}^{\mbox{\scriptsize{(0,0,+)}}}(\eta) \right)\right]\, .
\end{align*}
The kernels of the integral operators  $\hat{\textsf{V}}^{\mbox{\scriptsize{(0,0,$+ $)}}}$ and $\hat{\textsf{R}}^{\mbox{\scriptsize{(0,0,$+$)}}}$
are defined in (\ref{v0}) and (\ref{rp0}) and they act on $[-k_F,k_F]$  with $k_F=(h+|B|)^{1/2}$. The $\Gp{2}(x,t)$ correlator describes the
situation  in which a single particle of type 2 is injected in a sea of particles of type 1 (see also \cite{GPZ1,GPZ2}).

\subsection{Single component limit}\label{scl}

In the limit $B\rightarrow -\infty$, $h\rightarrow\-\infty$ such that $h_1=h-B$ is finite the energy
of the second type of particles becomes infinite ($h+B\rightarrow\infty)$ and they are effectively excluded
from the system. In this limit the model reduces to a single component system with chemical
potential $h_1$ and $\gamma=1$ for which $F(\gamma=1,\eta)=2\pi\delta(\eta)$. The Fermi function becomes
$\vartheta(k)=(1+e^{(k^2-h_1)/T})^{-1}$  and the correlators are
\begin{align}\label{1minus}
\left.\Gm{1}(x,t)\right|_{\overset{h,B\rightarrow-\infty}{h_1=h-B}}&=e^{- i th_1}
\left[\mbox{\textsf{det}}\left(1+ \hat{\textsf{V}}^{\mbox{\scriptsize{(T,--)}}}(0)+
\, \hat{\textsf{R}}^{\mbox{\scriptsize{(T,--)}}}\right)
- \mbox{\textsf{det}}\left(1+ \hat{\textsf{V}}^{\mbox{\scriptsize{(T,--)}}}(0) \right)\right]\, ,\\
\left.\Gp{1}(x,t)\right|_{\overset{h,B\rightarrow-\infty}{h_1=h-B}}&=e^{+ i t h_1}
\left[\mbox{\textsf{det}}\left(1+ \hat{\textsf{V}}^{\mbox{\scriptsize{(T,+)}}}(0)-
\hat{\textsf{R}}^{\mbox{\scriptsize{(T,+)}}}(0)\right)
+(G(x,t)-1)\,  \mbox{\textsf{det}}\left(1+ \hat{\textsf{V}}^{\mbox{\scriptsize{(T,+)}}}(0) \right)\right]\, .
\end{align}
The integral operators act on the real axis and the kernels are
\begin{align}\label{1plus}
\textsf{V}^{\mbox{\scriptsize{(T,$\pm $)}}}(0,\kappa\, |\, k,k')&=4 \sin^2\left( \frac{\pi \kappa}{2}\right)\, \sqrt {\vartheta(k)}\,
\frac{E^{(\pm)}(k)\,e(k')-E^{(\pm)}( k')\,e(k)}{2\pi i (k-k')}\, \sqrt {\vartheta(k')}\, ,\ \ \ \nu^{(\pm)}=\mp\frac{\kappa}{2}\, ,\\
\textsf{R}^{\mbox{\scriptsize{(T,+)}}}(0,\kappa\, |\, k,k')&=-2 \sin^2\left( \frac{\pi \kappa}{2}\right)\, \sqrt {\vartheta(k)}\,
\frac{E^{(+)}(k)E^{(+)}(k')}{\pi} \, \sqrt {\vartheta(k')}\, ,\ \ \
\textsf{R}^{\mbox{\scriptsize{(T,--)}}}(0,\kappa\, |\, k,k')\, =\sqrt {\vartheta(k)}\,
\frac{e(k) e(k')}{2 \pi} \, \sqrt {\vartheta(k')}\, .
\end{align}
At $\kappa=1$ this result reduces to the determinant representation obtained by Korepin and Slavnov for impenetrable bosons \cite{KS,KBI}
and for arbitrary statistics  the representation derived for single component anyons in \cite{PKA3} (note that the
statistics parameter $\kappa'$ used in \cite{PKA3} is related to ours via $\kappa'=1+\kappa$).

\section{Large distance asymptotics of static correlators at $T=0$ and $B=0$}\label{s4}

The determinant representations presented in the previous section are important for several reasons. Not only
that they constitute the starting point in proving that the correlators of quantum integrable systems  are
governed by classical integrable systems \cite{IIKS1,IIK1,KBI} but they are also particularly suited for the derivation
of the asymptotic behavior both at small and large spatial separations. We will see that the asymptotic analysis
is closely connected with the solution of an associated Riemann-Hilbert problem. Employing this method rigorous
results can be derived which then can be compared with the TLL/CFT predictions or, as in the case of the
spin-incoherent Tomonoga-Luttinger liquid, they can reveal totally new regimes. In addition to being extremely
useful in deriving analytical results the determinant representations can also be efficiently implemented
numerically \cite{Born} providing accurate information on the correlators. Coupled with the results on the asymptotic
behavior the numerical evaluation of the determinants can be used to compute the momentum distribution or the
time and space Fourier transform of the dynamical correlators which are both experimentally accessible quantities.
In this section we are going to derive the large distance asymptotic behavior of the static correlators at $T=0$
and $B=0$. This is a very interesting regime because as we will see, compared with the single component system
which is characterized by an algebraic decay of the correlators, in our case we will have an exponential decay
even though we are at zero temperature. This is due to the fact that our system is in the spin-incoherent
regime. We are
also going to numerically evaluate the momentum distribution and show that is nonsymmetric for $\kappa\ne\{0,1\}$
and that it presents a universal $C/k^4$ tail with $C$ the Tan's contact.

The determinant representation for the static correlators in the ground state of the balanced system can be
obtained from (\ref{corrsm}) and (\ref{corrsp}) by noticing that if we take the $B\rightarrow 0$ and
$T\rightarrow 0$ limits successively we have $\gamma=2$ and $\vartheta(k)=\theta(k^2-k_F^2)/2$ with $k_F=h^{1/2}$.
From now on we will consider $x\ge 0$. Results for the $x<0$ case  can be derived using (\ref{staticsymm}).
For the $\Gm{1}(x,0)$ correlator we obtain
\be\label{corrsm0}
\Gm{1}(x,0)=
\mbox{\textsf{det}}\left(1+ \hat{\textsf{v}}^{\mbox{\scriptsize{(0,--)}}}+
\hat{\textsf{r}}^{\mbox{\scriptsize{(0,--)}}}\right)
- \mbox{\textsf{det}}\left(1+ \hat{\textsf{v}}^{\mbox{\scriptsize{(0,--)}}} \right)\, ,
\ee
and $\Gp{1}(x,0)=\delta(x)-e^{-i\pi\kappa}\Gm{1}(-x,0)$. The operators $\hat{\textsf{v}}^{\mbox{\scriptsize{(0,--)}}}$
and $\hat{\textsf{r}}^{\mbox{\scriptsize{(0,--)}}}$ act on $[-k_F,k_F]$ with kernels
\begin{align}\label{defvs0}
\textsf{v}^{\mbox{\scriptsize{(0,$- $)}}}(k,k')&= \xi \,
 \frac{\sin[(k-k') x/2)]}{\pi  (k-k')}\, ,\ \
\textsf{r}^{\mbox{\scriptsize{(0,$-$)}}}( k,k')=\frac{ e^{- i(k+k') x/2}}{ 4 \pi}
\, ,\ \ \ \xi=-\left(1-\frac{e^{i\pi\kappa}}{2}\right)\, .
\end{align}
Introducing two functions $e_\pm^s(k)$ these kernels can be written as
\begin{align}\label{i1}
\textsf{v}^{\mbox{\scriptsize{(0,$-$)}}}(k,k')&=\frac{\xi}{2 \pi i}
 \frac{e_+^s(k)e_-^s(k')-e_+^s(k')e_-^s(k)}{  k-k'}\, ,\ \
\textsf{r}^{\mbox{\scriptsize{(0,$-$)}}}( k,k')=\frac{ e_-^s(k)e_-^s(k')}{ 4 \pi}
\, ,\ \ \ e_{\pm}^s(k)=e^{\pm i k x/2}\, .
\end{align}
The particular factorization shown in (\ref{i1}) reveals that the  $\hat{\textsf{v}}^{\mbox{\scriptsize{(0,$-$)}}}$
operator is of a special type called integrable integral operators \cite{IIKS1,HI} for which the resolvent
belongs to the same algebra. More precisely, if we define $f_\pm(k)$ as the solutions of the integral equations
\begin{align}\label{deffs}
f_+(k)+\int_{-k_F}^{k_F} \textsf{v}^{\mbox{\scriptsize{(0,--)}}}(k,k') f_+(k')\, dk'=e_+^s(k)\, ,\ \
f_-(k)+\int_{-k_F}^{k_F} \textsf{v}^{\mbox{\scriptsize{(0,--)}}}(k,k') f_-(k')\, dk'&=e_-^s(k)\, ,
\end{align}
the resolvent operator which satisfies $(1-\hat{\textsf{w}}^{\mbox{\scriptsize{(0,--)}}})=
(1+\hat{\textsf{v}}^{\mbox{\scriptsize{(0,--)}}})^{-1}$ has a kernel given by the same formula as the one
for $\textsf{v}^{\mbox{\scriptsize{(0,$-$)}}}(k,k')$ given in (\ref{i1})  with the functions $e_\pm^s(k)$
replaced by $f_\pm(k)$.
%
%\begin{align}\label{i2}
%\textsf{w}^{\mbox{\scriptsize{(0,$-$)}}}(k,k')&=\, \mbox{sgn}(x)\, \frac{\xi}{2 \pi i}
% \frac{f_+(k)f_-(k')-f_+(k')f_-(k)}{ k-k'}\, .
%\end{align}
%
We also point out that $e_+^s(k)e_-^s(k)-e_+^s(k)e_-^s(k)=0$ which means that the kernel is nonsingular on the diagonal.

We want to rewrite (\ref{corrsm0}) in a form which is more amenable to extracting the asymptotic behavior
at large $x$. Noticing that $\textsf{r}^{\mbox{\scriptsize{(0,$-$)}}}( k,k')$  is a rank 1 matrix and
using the fact that for a linear function in $z$ we have $\frac{\6}{\6 z} f(z)|_{z=0}=f(1)-f(0)$ we obtain
\begin{align}\label{i3}
\Gm{1}(x,0)&=
\frac{\6}{\6 z}\left. \mbox{\textsf{det}}\left(1+ \hat{\textsf{v}}^{\mbox{\scriptsize{(0,--)}}}+z\, \hat{\textsf{r}}^{\mbox{\scriptsize{(0,--)}}}\right) \right|_{z=0}
=\mbox{Tr}\, \left[\left(1+ \hat{\textsf{v}}^{\mbox{\scriptsize{(0,--)}}}\right)^{-1} \hat{\textsf{r}}^{\mbox{\scriptsize{(0,--)}}}\right]
 \mbox{\textsf{det}}\left(1+ \hat{\textsf{v}}^{\mbox{\scriptsize{(0,--)}}}\right)\nonumber\, ,\\
 &=\frac{1}{4\pi} \int_{-k_F}^{k_F}\int_{-k_F}^{k_F} \left(1+ \hat{\textsf{v}}^{\mbox{\scriptsize{(0,--)}}}\right)^{-1}(k,k')\, e_-^s(k)\, e_-^s(k')\, dk\, dk'\,
 \mbox{\textsf{det}}\left(1+ \hat{\textsf{v}}^{\mbox{\scriptsize{(0,--)}}}\right)\, ,\nonumber\\
 &=\frac{1}{4\pi} B_{--}\, \mbox{\textsf{det}}\left(1+ \hat{\textsf{v}}^{\mbox{\scriptsize{(0,--)}}}\right)\, ,
\end{align}
where we have used (\ref{aa1}) in the first line, the definition of the $f_\pm(x)$ function (\ref{deffs})
in the second line and introduced the notation $B_{ab}=\int_{-k_F}^{k_F} e_a^s(k)\,  f_b(k)\, dk$ with
$ a,b=\pm$. The objects $B_{ab}$ are called potentials and they play an important role in the asymptotic
analysis. We  also need to introduce $\vec{e}(k)$ and $\vec{f}(k)$ defined by $\vec{e}(k)= \left(e_+^s(k)\, ,
e_-^s(k)\right)^T$ and  $\vec{f}(k)= \left(f_+(k)\, , f_-(k)\right)^T .$

\subsection{Riemann-Hilbert problem for the static corelator}

From (\ref{i3}) we see that in order to obtain the large distance asymptotic behaviour of the static correlator
we need to analyse the large $x$  behaviour of both $B_{--}(x)$ and the Fredholm determinant. Our task is considerably simplified by using
a very powerful result of  Kitanine, Kozlowski, Maillet, Slavnov and Terras \cite{KKMST} which investigated the
more complicated case of the  generalized sine-kernel of which our kernel is a particular case. In \cite{KKMST}
the authors were mainly interested in the determinant asymptotics but their analysis can also be used to derive
the large distance behaviour of $B_{--}(x)$.  The Riemann-Hilbert problem associated with our determinant is the
following (Prop. 3.1 of \cite{KKMST}). Let $\hat{\textsf{v}}^{\mbox{\scriptsize{(0,$-$)}}}$  be the integral
operator defined in (\ref{defvs0}) acting on $L^2([-k_F,k_F])$ and such that $\mbox{\textsf{det}}\left(1+
\hat{\textsf{v}}^{\mbox{\scriptsize{(0,--)}}}\right)\ne 0$. Then there exists a 2-by-2 matrix $\chi(k)$ such that
$\vec{f}(k)=\chi(k)\vec{e}(k)$ and is the unique solution of the RHP:

\begin{itemize}
\item $\chi(k)$ is analytic on $\mathbb{C}\backslash[-k_F,k_F]$.

\item $\chi_+(k)G_\chi(k)=\chi_-(k)$ for $k\in(-k_F,k_F)$ with jump matrix
\be
G_\chi(k)=\left(\begin{array}{cc}
                    1-\xi&\xi\, [{e_+^s}(k)]^2\\
                    -\xi\, [e_-^s(k)]^2& 1+\xi
                    \end{array}  \right)\, .
\ee
Here $\chi_{\pm}(k')$ represent the limiting values of $\chi_{\pm}(k)$ when $k$ tends to the point $k'$ of $(-k_F,k_F)$
from the left, respectively right, side of the contour.

\item  $\chi(k)\underset{k\rightarrow\infty}{\longrightarrow} I_2= \left(\begin{array}{cc}
                    1&0 \\
                    0&1
     \end{array}\right)\,   $
and $\chi(k)=O\left(\begin{array}{cc}
                    1&1 \\
                    1&1
                    \end{array}  \right) \ln |k^2-k_F^2|$ for $k\rightarrow\pm k_F$\, .

\item Also
\be
\chi(k)=I_2-\frac{\xi}{2\pi i}\int_{-k_F}^{k_F} \frac{1}{k'-k}  \left(\begin{array}{cc}
                   - e_-^s(k')f_+(k')& e_+^s(k')f_+(k') \\
                   - e_-^s(k')f_-(k')   &e_+^s(k')f_-(k')
    \end{array}\right)\, dk'\, .
\ee
\end{itemize}

The last relation is extremely important because it shows that the potentials $B_{ab}(k)$ can be obtained from the
large-$k$  expansion of the RHP solution
\be\label{i4}
\lim_{k\rightarrow\infty} \chi(k)=I_2+\frac{1}{k}\frac{\xi}{2\pi i} \left(\begin{array}{cc}
                    -B_{-+}  & B_{++} \\
                    -B_{--}  & B_{+-}
     \end{array}\right)\,+O\left(\frac{1}{k^2}\right)\, .
\ee
In general the derivative of the Fredholm determinant with respect to $x$ or $t$ (in the dynamical case) is expressed
in terms of the $B_{ab}(k)$ potentials. Therefore, the asymptotic behavior of the determinant can be derived from the
large $k$ and $x$ expansion of the RHP solution.

\subsection{Large distance asymptotic behavior of the static correlator}

\begin{figure*}
\includegraphics[width=0.75 \linewidth]{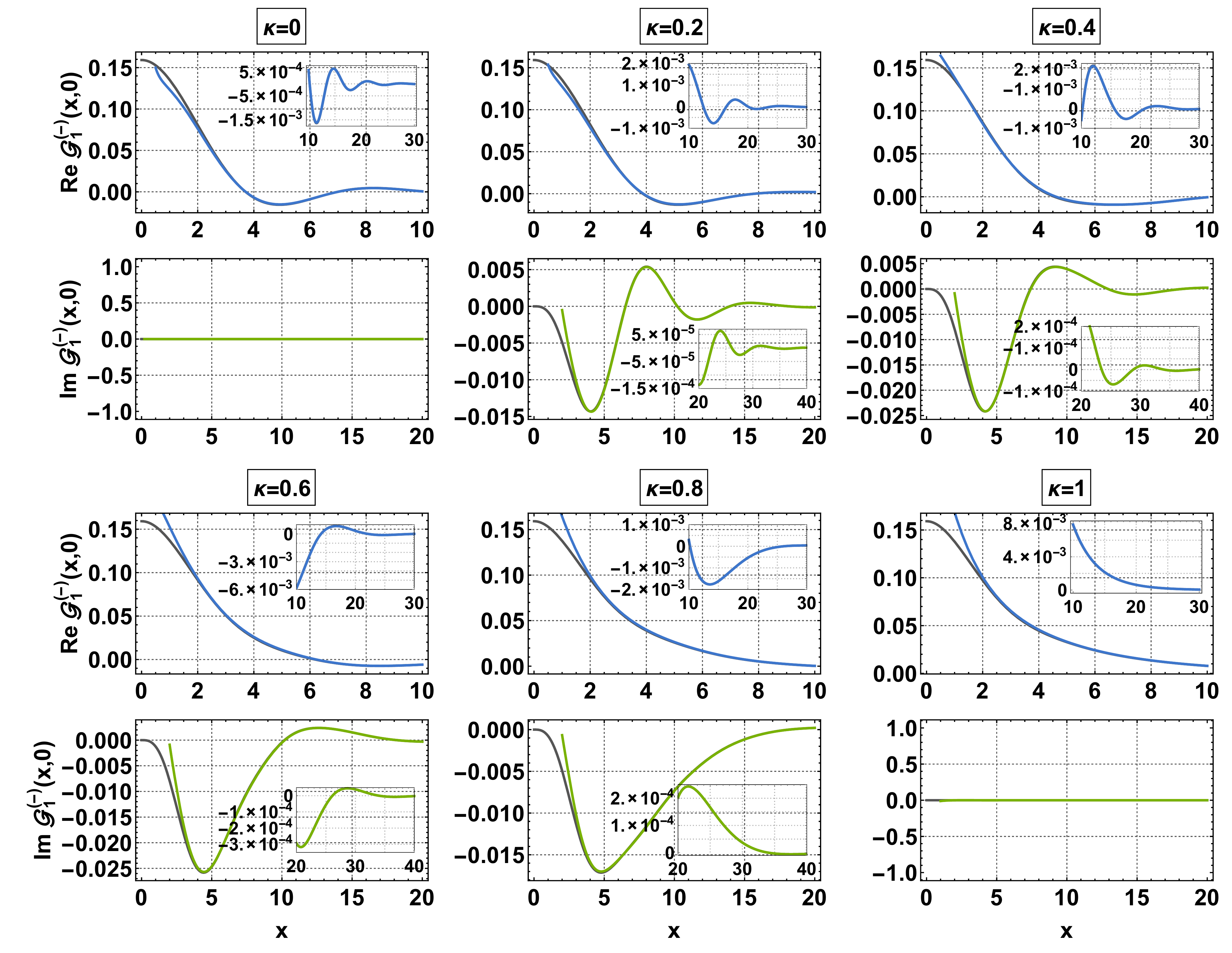}
\caption{Plots of the static correlator $\mathcal{G}_1^{(-)}(x,0)$ at $T=0$, $B=0$ and $h=1$ (black line) and the
asymptotic formula (\ref{asymptstatic}) (real part blue line, imaginary part green line) for various values of the
statistics parameter. }
\label{FigStatic}
\end{figure*}

The Fredholm determinant appearing in (\ref{corrsm0}) is a particular case of the generalized sine-kernel studied
in \cite{KKMST}. Introducing
\be\label{defnus}
\nu\equiv-\frac{1}{2\pi i}\ln(1+\xi)=-\frac{i \ln 2}{2\pi}-\frac{\kappa}{2}\, ,
\ee
the asymptotic behavior of the determinant for large $x$ is given by the following theorem:
\begin{thm}\cite{KKMST}\label{thm1}
Let $\hat{\textsf{v}}^{\mbox{\scriptsize{(0,$- $)}}}$ be the integral operator with the kernel defined in (\ref{defvs0}).
For $\kappa\in[0,1)$ and $x\rightarrow\infty$ we have
\be
 \mbox{\textsf{det}}\left(1+ \hat{\textsf{v}}^{\mbox{\scriptsize{(0,--)}}}\right)=\frac{G^2_B(1,\nu)}{(2k_F)^{2\nu^2}}\,  \frac{e^{-2i k_F \nu x}}{  x^{2\nu^2}}
\, ,
\ee
where $G_B(1,z)=G_B(1+z)G_B(1-z)$ and $G_B(z)$ is Barnes G-function \cite{Barnes} which has the integral representation
\be
G_B(z+1)=(2\pi)^{z/2}\exp\left\{-\frac{z(z-1)}{2}+\int_0^z t\Psi(t)\, dt\right\}\, ,\ \ \mbox{Re}(z)>-1\, ,
\ \ \Psi(z)=\frac{\Gamma'(z)}{\Gamma(z)}\, .
\ee
\end{thm}
We should point out that the analysis in \cite{KKMST} is valid only for $\kappa\in[0,1)$  so in principle this result
cannot be used in the case of two component bosons $(\kappa=1)$. Nevertheless numerical data
obtained from the evaluation of the  representation (\ref{corrsm0}) agrees with the asymptotic behavior derived by
taking the limit $\kappa=1$ in the previous result. Thm.~\ref{thm1} together with the expression for $B_{--}$
(\ref{aa5}) which is derived in Appendix ~\ref{a2} allows us to state the main result of this section, the large distance
asymptotic behavior of the static correlators at zero temperature and zero magnetic field is
\be\label{asymptstatic}
\Gm{1}(x,0)=\frac{\pi e^{\mathcal{C}(\nu)}}{2\xi\sin\pi\nu}e^{-2 i k_F \nu x}x^{-(2\nu^2+1)}
\left[\frac{(2k_F x)^{-2\nu}}{\Gamma^2(-\nu)} e^{-i k_F x}-\frac{(2k_F x)^{2\nu}}{\Gamma^2(\nu)} e^{i k_F x}\right]\, ,
\ee
with $\nu$ defined in (\ref{defnus}),  $\xi$ in (\ref{defvs0}) and
\be\label{defconst}
\mathcal{C}(\nu)=-2\nu^2\left[1+\ln(2k_F)\right]+2\nu\ln\left[\frac{\Gamma(\nu)}{\Gamma(-\nu)}\right]-2\int_0^\nu
\ln\left[\frac{\Gamma(t)}{\Gamma(-t)}\right]\, dt\,
=\ln\left[\frac{G_B^2(1,\nu)}{(2k_F)^{2\nu^2}}\right]\, .
\ee
This result is valid for large $x>0$. For $x<0$ we use $\mathcal{G}_1^{(-)}(x,0)=\overline{\mathcal{G}}_1^{(-)}(-x,0).$
In Fig.~\ref{FigStatic} we present the numerical evaluation of the static correlator and our asymptotic formula
(\ref{asymptstatic}) and we see that already for $x>5$ the two curves become indistinguishable. The accuracy of the expansion
is the largest for $\kappa=0$ (fermions) and decreases as we increase the statistics parameter but it is still very accurate
even in the bosonic case.

In the fermionic case the asymptotic formula (\ref{asymptstatic}) was first obtained by Berkovich and Lowenstein \cite{BL,B1}
(a different proof similar with the one presented here was later derived in \cite{CZ2}). For $\kappa=0$ (\ref{asymptstatic})
can be written as
\be
\Gm{1}(x,0)\overset{\kappa=0}{=}\mathcal{C}'(0)e^{-\frac{\ln 2}{\pi} k_F x} x^{-1+\frac{1}{2}\left(\frac{\ln 2}{\pi}\right)^2}
\sin\left(k_F x-\ln 2\ln x/\pi-\varphi_0\right)\, ,
\ee
with $\mathcal{C}'(0)=-4\pi\sqrt{2}\exp\{\mathcal{C}[- i\ln2/(2\pi)]-2\, \mbox{Re}\left[\ln \Gamma(i\ln2/2\pi)\right]\}$ and
$\varphi_0=\ln 2\ln(2k_F)/\pi-2\,\mbox{Im}\left[\ln \Gamma(i\ln 2 /2\pi)\right]\, .$ In the bosonic case $(\kappa=1)$  the
first term in the square parenthesis of (\ref{asymptstatic}) gives the leading contribution and we find
\be
\mathcal{G}_1^{(-)}(x,0)\overset{\kappa=1}{=}-\left(\frac{\pi e^{\mathcal{C}(\nu)}}{3\sin\pi\nu}\right)\frac{(2k_F)^{-2\nu}}{\Gamma^2(-\nu)}
e^{-\frac{\ln 2}{\pi} k_F x}x^{-\frac{1}{2}+\frac{1}{2}\left(\frac{\ln 2}{\pi}\right)^2}\, ,\ \ \nu=-\frac{i \ln 2}{2\pi}-\frac{1}{2}\, .
\ee
This result appeared first, without the constant and derivation, in \cite{CSZ} (please note that their result corresponds to $k_F=2$).

The most striking feature of the asymptotic expansion for the static correlators is the exponential decay  even though
we are at zero temperature. This is a consequence of the fact that we are considering the correlators in the spin-incoherent
regime which is characterized by a highly excited spin sector. A computation of the entropy per length from Takahashi's
formula (\ref{takahashi}) gives  $s=\ln 2$ even at zero temperature which shows that the spin sector is completely disordered.
The exponential decay term is the same  for all values of the statistics parameter, $e^{-\frac{\ln 2}{\pi} k_F x}$ and it depends
only on the number of components of the system. This is another particularity of the spin-incoherent system \cite{F}, for an
$N$-component system  the exponential decay is given by $e^{-\frac{\ln N}{\pi} k_F x}$. This exponential decay is accompanied
by an oscillatory  term with frequency proportional with the statistics $e^{i\kappa k_F x}.$ The algebraic corrections become
smaller  as we increase $\kappa$. In the vicinity of the fermionic point, $\kappa=0$, both terms of the expansion in \ref{asymptstatic}
are important while for  $\kappa\sim 1$  only one term gives the leading contribution. This mirrors a similar behavior of the
correlators of single component impenetrable anyons for which the algebraic decay at zero temperature is given by $(x>0)$ \cite{CM}
\be\label{1CFT}
\mathcal{G}_1^{(-)}(x,0)\sim b_0 \frac{e^{i(1-\kappa)k_F x}}{x^{1-\kappa+\kappa^2/2}}+ b_{-1} \frac{e^{-i(1+\kappa)k_F x}}{x^{1+\kappa+\kappa^2/2}}\, ,
\ee
with $b_0, b_{-1}$ constants.

\subsection{Momentum distribution and contact}

\begin{figure*}
\includegraphics[width=0.75 \linewidth]{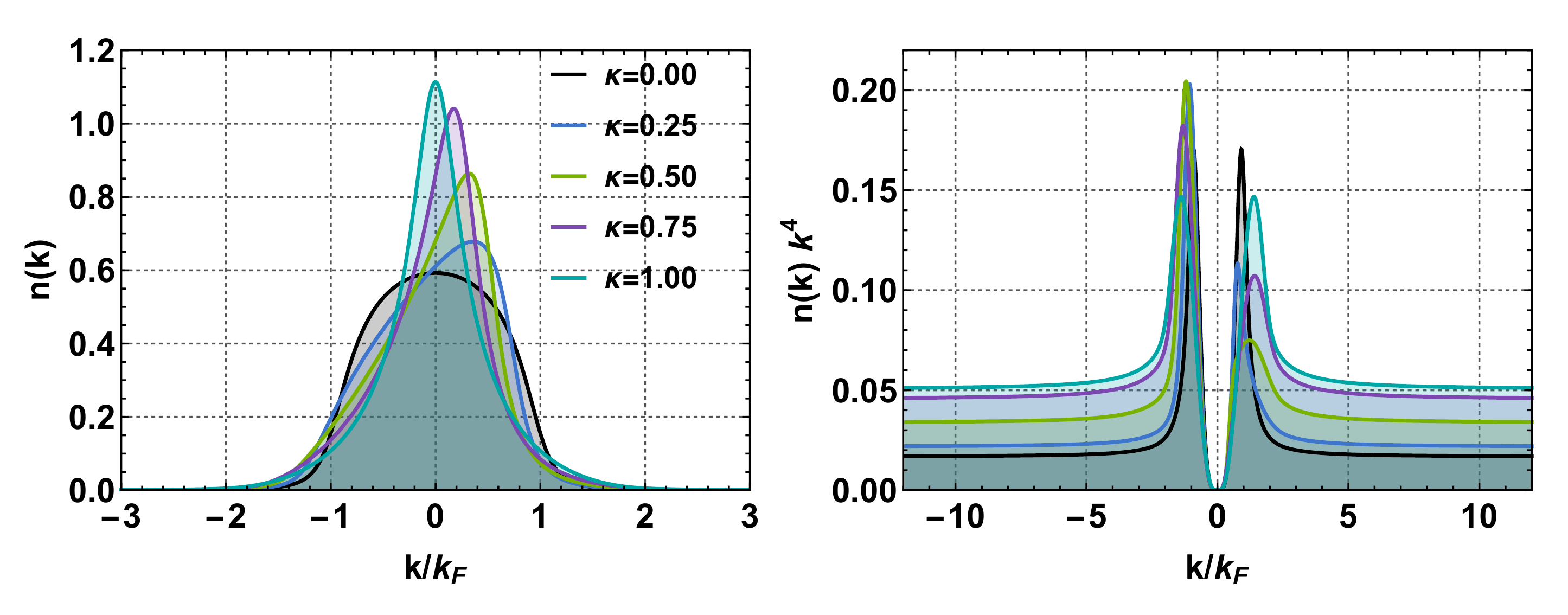}
\caption{(Left panel) Momentum distribution $n(k)$ at $T=0, B=0$  and $h=1$  for various values of the statistics parameter.
(Right panel) Plot of $n(k)\, k^4$  showing the $1/k^4$ behavior of the momentum distribution at large momenta. The asymptotic
value at large $|k|$ for each value of $\kappa$ is the contact. }
\label{FigMomentum}
\end{figure*}

The experimentally relevant  momentum distribution is defined as the Fourier transform of the static correlator $n_1(k)=
\int e^{- i k x} \mathcal{G}_1^{(-)}(x,0)\, dx\, .$ While in the balanced system we have $n(k)=n_1(k)=n_2(k)$, in the presence
of a magnetic field the momentum distributions of each type of particles is different. A simple consequence of
$\mathcal{G}_1^{(-)}(x,0)=\overline{\mathcal{G}}_1^{(-)}(-x,0)$ is that
\be
n(k)=2\int_0^\infty \cos (k x)\, \mbox{Re}\, \mathcal{G}_1^{(-)}(x,0)\, dx+ 2\int_0^\infty \sin(k x)\,
\mbox{Im}\, \mathcal{G}_1^{(-)}(x,0)\, dx\, .
\ee
In the fermionic and bosonic case $\mbox{Im}\, \mathcal{G}_1^{(-)}(x,0)=0$ and therefore the momentum distribution is symmetric,
$n(k)=n(-k)$, but for an arbitrary value of the statistics parameter the static correlator has a nonzero imaginary part which
results in a nonsymmetric momentum distribution as it can  be seen in the left panel Fig.~\ref{FigMomentum}. Compared with the
single component case $n(k)$ does not have any singularities due to the exponential decay. An universal property of models with
contact interactions is that the tail of the momentum distribution is given by $\lim_{|k|\rightarrow\infty}n(k)=C/k^4$ with $C$
a quantity called contact \cite{Tan1,Tan2,Tan3,OD,BP1,ZL,WC1,WC2,VZM1,VZM2,BZ,PK2}. This is due to the discontinuity of the
wavefunction's derivative at the coinciding points and, as  it can be seen in the right panel of Fig.~\ref{FigMomentum},  even
though the momentum distribution is nonsymmetric the tails  exhibit the $1/k^4$ behavior with the amplitude given  by the contact.
For two-component impenetrable anyons the contact is a  monotonic function of the statistics parameter with the minimum obtained for the
fermionic system and reaching its maximum for the bosonic system.

\section{Long time large distance asymptotics of dynamic correlators at $T=0$ and $B=0$}\label{s5}

The derivation of the asymptotic behavior of time and space dependent correlators is more involved than
the static case but it follows along the same lines. The Fredholm determinants appearing in the representations
for the dynamic correlators are also integrable which means that their asymptotics can be derived by
solving an associated RHP. Their kernels are particular cases of the time dependent generalization of the sine-kernel
for which a comprehensive asymptotic analysis was performed in \cite{K2} by Kozlowski.
We are interested in the large $x$ behavior of the
dynamic correlators for a fixed value of the saddle point  $k_0=x/2t$. The results will be different
depending on the position of the saddle point with respect to the Fermi points $\pm k_F$. If  $|x|>2k_F|t|$
we will say that we are in the space-like regime and for  $|x|<2k_F |t|$ we are in the time-like regime.
Our results are not valid for $k_0=\pm k_F$.

At zero temperature and no magnetic field the determinant representation for dynamic correlators is given by (\ref{corr00}) with the
kernels defined in (\ref{kernels00}). The operators $\hat{\textsf{V}}^{\mbox{\scriptsize{(0,0,$\pm$)}}}$ are
integrable with the kernels of the resolvents which satisfy $\left(1-\hat{\textsf{W}}^{\mbox{\scriptsize{(0,0,$\pm$)}}}
\right)=\left(1+\hat{\textsf{V}}^{\mbox{\scriptsize{(0,0,$\pm$)}}}\right)^{-1}$ given by
\be
\textsf{W}^{\mbox{\scriptsize{(0,0,$\pm $)}}}(\eta,\kappa\, |\, k,k')=4 \sin^2 \pi \nu^{(\pm)}\,
\frac{F^{(\pm)}(k)\,f(k')-F^{(\pm)}( k')\,f(k)}{2\pi i (k-k')}\, ,\ \ \ \nu^{(\pm)}=\pm\left(\frac{\eta}{2\pi}-\frac{\kappa}{2}\right)\, ,\\
\ee
where the functions $F^{(\pm)}(k)$ and $f(k)$ are solutions of the integral equations
\begin{align}
F^{(\pm)}(k)+\int_{-k_F}^{k_F} \textsf{V}^{\mbox{\scriptsize{(0,0,$\pm$)}}}(k,k') F^{(\pm)}(k')\, dk'&=E^{(\pm)}(k)\, ,\ \
f(k)+\int_{-k_F}^{k_F} \textsf{V}^{\mbox{\scriptsize{(0,0,$\pm$)}}}(k,k') f(k')\, dk'=e(k)\, .
\end{align}
%
%$\vec{E}(k)=\left(E^{(\pm)}(k),e(k)\right)^T$ and $\vec{F}(k)=\left(F^{(\pm)}(k),f(k)\right)^T$ %
Introducing the potentials
$ B_{++}=\int_{-k_F}^{k_F} E^{(\pm)}(k) F^{(\pm)}(k)\, dk\, ,$
$ B_{--}=\int_{-k_F}^{k_F} e(k)f(k)\, dk\, ,$
$ B_{+-}=\int_{-k_F}^{k_F} E^{(\pm)}(k) f(k)\, dk\, ,$
$ B_{-+}=\int_{-k_F}^{k_F} e(k) F^{(\pm)}(k)\, dk\, $
and
\be
H(\eta)\equiv\frac{3}{5-4\cos\eta}=1+\frac{e^{i\eta}}{2-e^{i\eta}}+\frac{e^{-i\eta}}{2-e^{-i\eta}}\, ,\ \
b_{++}=\frac{2\sin^2\pi\nu^{(\pm)}}{\pi}B_{++}+G(x,t)\, ,
\ee
then, analogous with the derivation of (\ref{i3}), we can derive a more suitable representation for the correlators
\begin{subequations}\label{corr002}
\begin{align}
\Gm{1}(x,t)&=e^{- i th}\int_{-\pi}^\pi \frac{d\eta}{2\pi}\frac{ H(\eta)}{4\pi}\, B_{--}(\eta)\,
\mbox{\textsf{det}}\left(1+ \hat{\textsf{V}}^{\mbox{\scriptsize{(0,0,--)}}}(\eta) \right)\, ,\label{detm02}\\
\Gp{1}(x,t)&=e^{+ i th}\int_{-\pi}^\pi \frac{d\eta}{2\pi}\ H(\eta)\,b_{++}(\eta)\,
 \mbox{\textsf{det}}\left(1+ \hat{\textsf{V}}^{\mbox{\scriptsize{(0,0,+)}}}(\eta) \right)\, .\label{detp02}
\end{align}
\end{subequations}
Making the change of variable $z=e^{i\eta}$ we obtain integrals over the unit circle with $H(z)=1+z/(2-z)+1/(2 z-1),$
\begin{subequations}
\begin{align}
\Gm{1}(x,t)&=e^{- i th}\oint _{|z|=1} \frac{dz}{2\pi}\frac{ H(z)}{4 i\pi z}\, B_{--}(z)\,
\mbox{\textsf{det}}\left(1+ \hat{\textsf{V}}^{\mbox{\scriptsize{(0,0,--)}}}(z) \right)\, ,\label{i44}\\
\Gp{1}(x,t)&=e^{+ i th}\oint _{|z|=1}\frac{dz}{2\pi}\ \frac{ H(z)}{iz}\,b_{++}(z)\,
 \mbox{\textsf{det}}\left(1+ \hat{\textsf{V}}^{\mbox{\scriptsize{(0,0,+)}}}(z) \right)\label{i5}\, .
\end{align}
\end{subequations}
The function $H(z)$ has two poles in the complex plane situated at $z_1=1/2$ and $z_2=2$ with residues $Res\, H(z_1)=1/2$ and
$Res\, H(z_2)=2$. Next, we are going to assume that the integrand on the right hand side of (\ref{i44}) (excluding $H(z)/z$) is analytic
in the annulus $1\le z<e$ and the integrand on the right hand side of (\ref{i5}) (also excluding $H(z)/z$) is analytic in the
annulus $0< z\le 1.$ Under these assumptions we can deform the contour in (\ref{i44}) to a circle of radius $r^{(-)}=2+a$ with
$a$ a small positive quantity and the contour in (\ref{i5}) to a circle of radius $r^{(+)}=1/2-a$  obtaining
\begin{subequations}
\begin{align}
\Gm{1}(x,t)&=
\frac{ e^{- i th}}{4 \pi}\, B_{--}(2)\,
\mbox{\textsf{det}}\left(1+ \hat{\textsf{V}}^{\mbox{\scriptsize{(0,0,--)}}}(2) \right)+
e^{- i th}\oint _{|z|=2+a} \frac{dz}{2\pi}\frac{ H(z)}{4 i\pi z}\, B_{--}(z)\,
\mbox{\textsf{det}}\left(1+ \hat{\textsf{V}}^{\mbox{\scriptsize{(0,0,--)}}}(z) \right)\, ,\label{i6}\\
\Gp{1}(x,t)&=
e^{+ i th}b_{++}(1/2)\,
 \mbox{\textsf{det}}\left(1+ \hat{\textsf{V}}^{\mbox{\scriptsize{(0,0,+)}}}(1/2) \right)
+e^{+ i th}\oint _{|z|=1/2-a}\frac{dz}{2\pi}\ \frac{ H(z)}{iz}\,b_{++}(z)\,
 \mbox{\textsf{det}}\left(1+ \hat{\textsf{V}}^{\mbox{\scriptsize{(0,0,+)}}}(z) \right)\label{i7}\, ,
\end{align}
\end{subequations}
where the first terms are given by the contribution of the residues of the function $H(z)$ at $z_1$  and $z_2$
(note that the contribution of the residue at $z_1$ comes with a minus sign due to the fact that the deformed
contour surrounds the pole clockwise). We will show below that the second terms give a negligible contribution in the
large $x$ limit so we can focus only on the first terms. Because $z=e^{i\eta}$ and $\nu^{(\pm)}=\pm\left(\frac{\eta}
{2\pi}-\frac{\kappa}{2}\right)$ this means that $\nu^{(+)}(z=1/2)=\left(\frac{i\ln 2}{2\pi}-\frac{\kappa}{2}\right)$ and
$\nu^{(-)}(z=2)=\left(\frac{i\ln 2}{2\pi}+\frac{\kappa}{2}\right).$

\subsection{RHP for dynamic correlators}

The RHP associated with the Fredholm determinants appearing in the representations for the dynamic
correlators is the following \cite{K2}:

\begin{itemize}
\item $\chi(k)$ is analytic on $\mathbb{C}\backslash[-k_F,k_F]$.

\item $\chi_+(k)G_\chi(k)=\chi_-(k)$ for $k\in(-k_F,k_F)$ with jump matrix
\be
G_\chi(k)=\left(\begin{array}{cc}
                    1-4 \sin^2\pi\nu^{(\pm)}E^{(\pm)}(k)e(k) & 4\sin^2\pi\nu^{(\pm)} (E^{(\pm)}(k))^2\\
                    - e^2(k)&  1+4 \sin^2\pi\nu^{(\pm)}E^{(\pm)}(k)e(k)
                    \end{array}  \right)\, .
\ee

\item  $\chi(k)\underset{k\rightarrow\infty}{\longrightarrow} I_2= \left(\begin{array}{cc}
                    1&0 \\
                    0&1
     \end{array}\right)\,    $ and
 $\chi(k)=O\left(\begin{array}{cc}
  1&1 \\
  1&1
  \end{array}  \right) \ln |k^2-k_F^2|$ for $k\rightarrow\pm k_F$\, .

\item Also
\be
\chi(k)=I_2-\frac{2\sin^2\pi\nu^{(\pm)}}{\pi i}\int_{-k_F}^{k_F} \frac{1}{k'-k}  \left(\begin{array}{cc}
                    - e(k')F^{(\pm)}(k')& E^{(\pm)}(k')F^{(\pm)}(k') \\
                    - e(k')f(k')   &E^{(\pm)}(k')f(k')
     \end{array}\right)\, dk'\, .
\ee
\end{itemize}

The last relation shows that
\be\label{i9}
\lim_{k\rightarrow\infty} \chi(k)=I_2+\frac{1}{k}\frac{2\sin^2\pi\nu^{(\pm)}}{\pi i} \left(\begin{array}{cc}
                    -B_{-+}  & B_{++} \\
                    -B_{--}  & B_{+-}
     \end{array}\right)\,+O\left(\frac{1}{k^2}\right)\, .
\ee

\subsection{Long-time, large-distance asymptotics}

\begin{figure*}
\includegraphics[width=0.73 \linewidth]{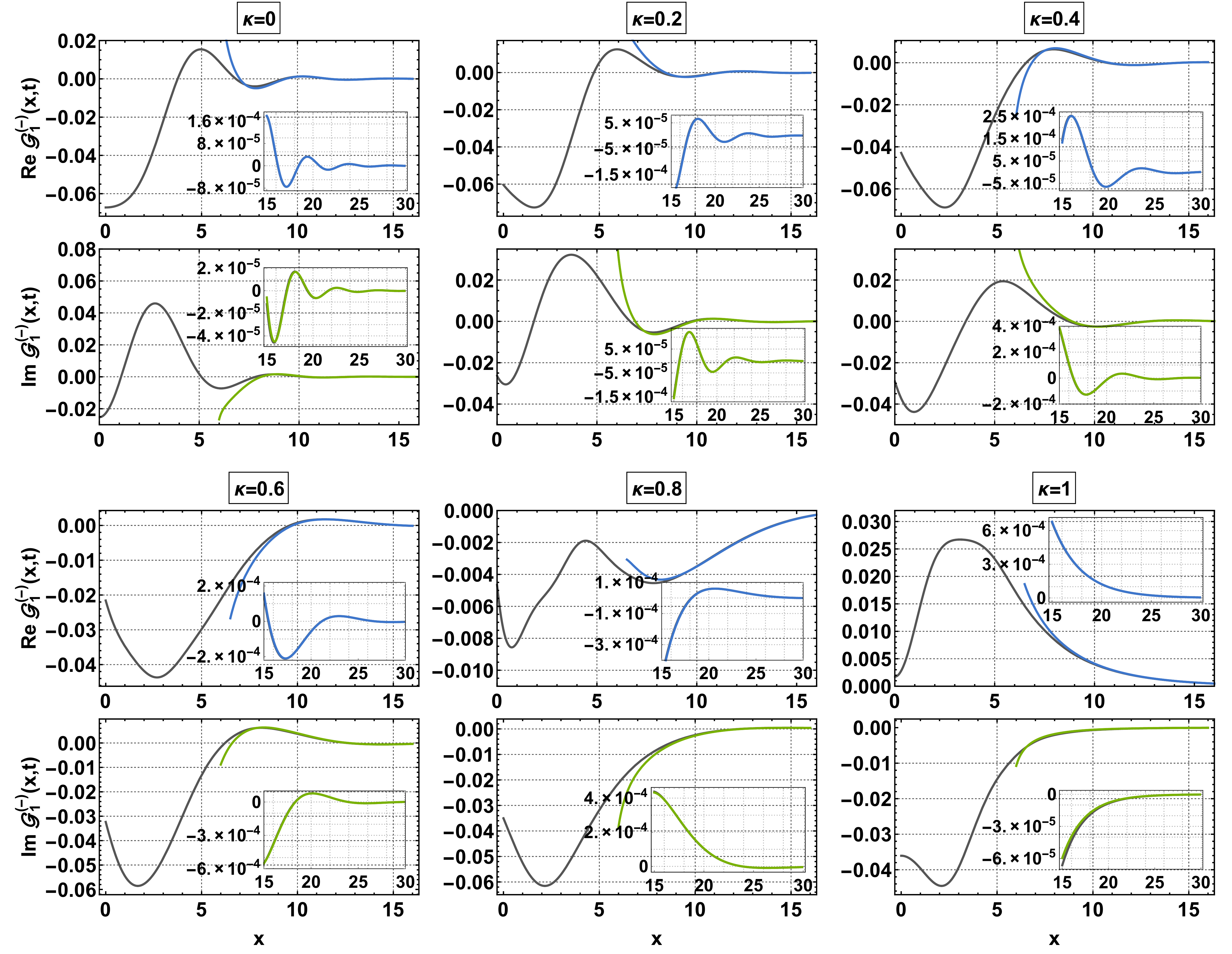}
\caption{Plots of the dynamic correlator $\mathcal{G}_1^{(-)}(x,t)$  at fixed $t=2$ and $h=2$ (black line) and the asymptotic formula
(\ref{gminusspace}) (real part blue line, imaginary part green line) for the space-like region. }
\label{gmspace}
\end{figure*}
\begin{figure*}
\includegraphics[width=0.73 \linewidth]{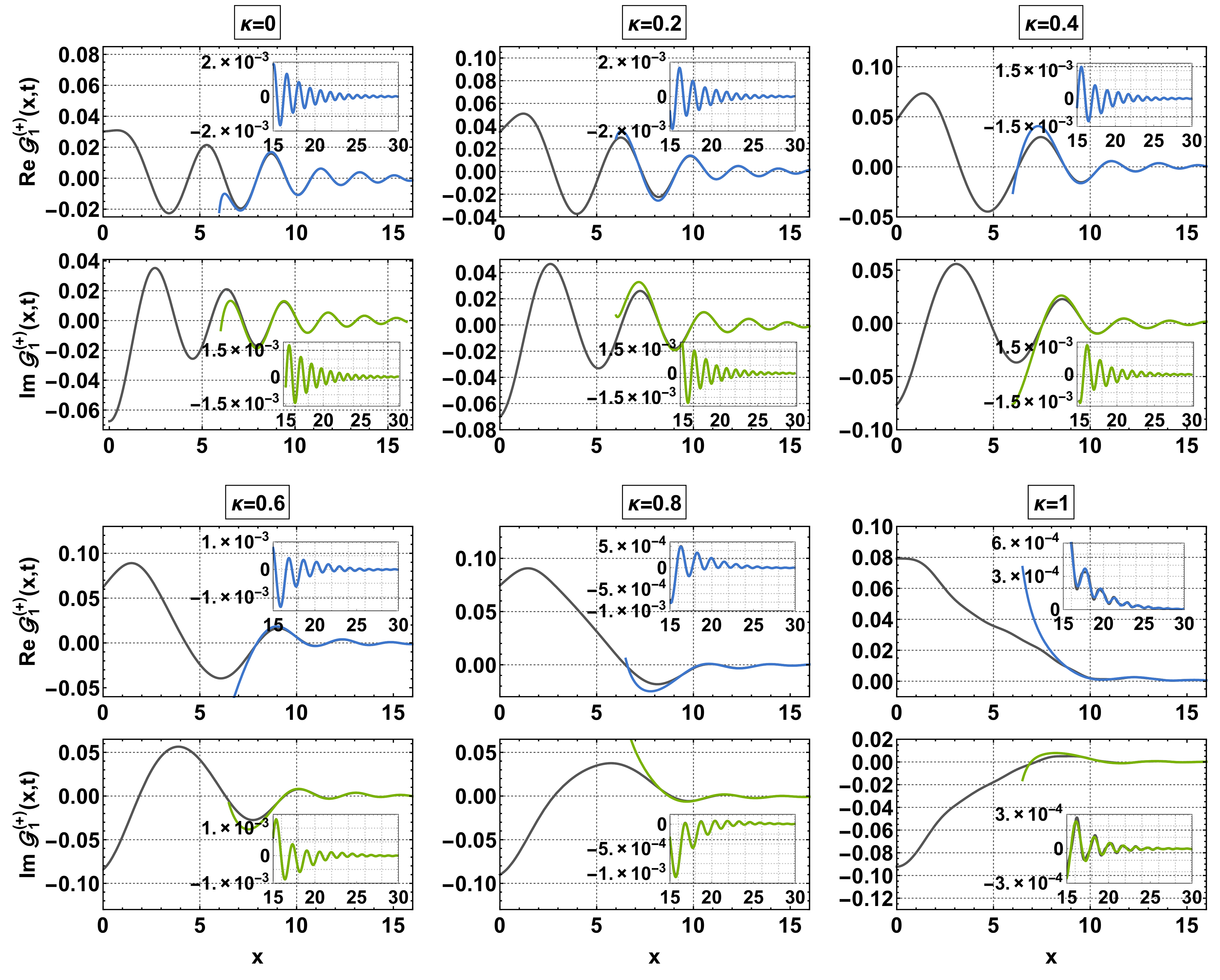}
\caption{Plots of the dynamic correlator $\mathcal{G}_1^{(+)}(x,t)$  at fixed $t=2$ and $h=2$ (black line) and the asymptotic formula
(\ref{gplusspace}) (real part blue line, imaginary part green line) for the space-like region. }
\label{gpspace}
\end{figure*}

The asymptotic solution of the RHP in  both space-like and time-like regime can be found in \cite{K2} and is briefly presented in
Appendices \ref{a3} and \ref{a4}. In this section we assume $x>0$ and $t>0$.
The asymptotic behavior of the time dependent Fredholm determinant is given by the following theorem:
\begin{thm}\label{thmd}\cite{K2}
Let $\hat{\textsf{V}}^{\mbox{\scriptsize{(0,0,$\pm$)}}}$ be an integral operator with the kernel defined in (\ref{kernels00})
and acting on $L^2([-k_F,k_F]).$ Then for $x\rightarrow\infty$ the leading asymptotic behavior is
\be
\mbox{\textsf{det}}\left(1+\hat{\textsf{V}}^{\mbox{\scriptsize{(0,0,$\pm$)}}}(\eta)\right)=\frac{G_B^2(1,\nu^{(\pm)})}{(2k_F)^{2(\nu^{(\pm)})^2}}
e^{2 i k_F x\nu^{(\pm)}} \frac{1}{\left(x-2 k_F t+i0^+\right)^{(\nu^{(\pm)})^2}}\frac{1}{\left(x+2 k_F t\right)^{(\nu^{(\pm)})^2}}\, ,
\ee
with $\nu^{(\pm)}=\pm\left(\frac{\eta}{2\pi}-\frac{\kappa}{2}\right)$ provided that $|\mbox{Re}\, \nu^{(\pm)}|<1/2.$
\end{thm}
The $i0^+$ regularization plays a role only in the time-like regime $(x<2k_F t)$.

\textit{Results in the space-like regime}. In the case of  $\mathcal{G}^{(-)}_1(x,t)$ we are interested in the
value of  the determinant and the $B_{--}$ potential evaluated at $z=2$ which corresponds to $\nu=\frac{i\ln 2}{2\pi}+\frac{\kappa}{2}$
(from now on we drop the $\pm$ superscripts when there is no potential for confusion).
The result for $B_{--}$ derived in Appendix \ref{a3} and Thm.~\ref{thmd}  show that the leading contribution in the second term on the right
hand side of (\ref{i6}) is given by the determinant which on the contour of integration is of the order of $\exp[-2k_F x \ln(2+a)/2\pi]$
and therefore is negligible  in the large $x$ limit. Using (\ref{i6}), (\ref{Bmmspace}) and Thm.~\ref{thmd} we obtain the
asymptotic behavior
\begin{align}\label{gminusspace}
\mathcal{G}^{(-)}_1(x,t)=e^{\mathcal{C}(\nu)}\frac{\pi}{8}\frac{\left(e^{-2i\pi\nu}-1\right)}{\sin^3\pi\nu}&e^{2ik_F\nu x}
\left[
\frac{(2k_F)^{-2\nu}}{\Gamma^2(-\nu)}\frac{e^{i k_F x}}{(x-2k_F t)^{\nu^2}(x+2k_F t)^{(\nu+1)^2}}\right.\nonumber\\
&\qquad\qquad\qquad\qquad\qquad\qquad\qquad\left.
-\frac{(2k_F)^{2\nu}}{\Gamma^2(\nu)}\frac{e^{-i k_F x}}{(x-2k_F t)^{(\nu-1)^2}(x+2k_F t)^{\nu^2}}
\right]\, ,
\end{align}
with $\nu=\frac{i\ln 2}{2\pi}+\frac{\kappa}{2}$ and $\mathcal{C}(\nu)$ defined in (\ref{defconst}).
For the $\mathcal{G}^{(+)}_1(x,t)$ correlator we need to evaluate the determinant and the $b_{++}$  potential
at $z=1/2$ which corresponds to $\nu=\frac{i\ln 2}{2\pi}-\frac{\kappa}{2}.$ The second term on the right hand side
of (\ref{i7}) is of the order of $\exp[2k_F x \ln(1/2-a)/2\pi]$ and therefore this contribution can be neglected.
Collecting the results from (\ref{i7}), (\ref{bppspace}) and Thm.~\ref{thmd} we find
\begin{flalign}\label{gplusspace}
\mathcal{G}^{(+)}_1(x,t)&=
e^{\mathcal{C}(\nu)}e^{2i k_F x\nu}\left\{\frac{e^{i th}G(x,t)}{(x-2k_F t)^{\nu^2}(x+2k_F t)^{\nu^2}}
\left(\frac{x-2k_F t}{x+2k_F t}\right)^{2\nu}\, \right.
\qquad\qquad\qquad\qquad\qquad\qquad\qquad\qquad\qquad\qquad\qquad\nonumber\\
& -\frac{\pi \left(e^{-2i\pi\nu}-1\right)^{-1}}{\sin\pi\nu}
\left.\left[
\frac{(2k_F)^{2\nu}}{\Gamma^2(\nu)}\frac{e^{-i k_F x}}{(x-2k_F t)^{\nu^2}(x+2k_F t)^{(\nu-1)^2}}
-\frac{(2k_F)^{-2\nu}}{\Gamma^2(-\nu)}\frac{e^{i k_F x}}{(x-2k_F t)^{(\nu+1)^2}(x+2k_F t)^{\nu^2}}
\right]\right\}\, .
\end{flalign}
with $\nu=\frac{i\ln 2}{2\pi}-\frac{\kappa}{2}.$
The asymptotic formulae (\ref{gminusspace}) and (\ref{gplusspace}) for the fermionic system were first derived by
Cheianov and Zvonarev \cite{CZ1,CZ2} who also pointed out the non-trivial, from the conformal point of view, behavior
of the correlation functions. The correlators show spin-charge separation with scaling behavior of the charge component and
exponential decay of the spin part. The anomalous dimensions of the charge part do not correspond to any unitary conformal
field theory. For arbitrary statistics the overall picture remains almost the same: the spin part is exponentially decaying in
space separation while the charge part presents scaling. In the fermionic case both terms in the square parenthesis are
important but as we increase the statistics parameter  one of them become dominant. Similar to the static case, in principle
(\ref{gminusspace}) and (\ref{gplusspace}) should not be valid in the bosonic case due to the fact that for $\kappa=1$ we have  $|\mbox{Re}\, \nu|=1/2$
which means Thm.~\ref{thmd} and the results of \cite{K2} cannot be used. However, by  taking the $\kappa\rightarrow 1$ limit in
the above asymptotic expansions we obtain results which agree with the numerics even in the bosonic case but we should point out that  the accuracy  is
decreasing as we approach $\kappa\rightarrow 1$.  In  Figs.~\ref{gmspace}  and \ref{gpspace} we plot the corelators for $h=2$ and fixed $t=2$ and
the expansions (\ref{gminusspace}) and (\ref{gplusspace}). For $x>2 k_F t$  we have almost perfect agreement if we take into consideration
that we have considered only the first terms of the expansion.

\begin{figure*}
\includegraphics[width=0.71 \linewidth]{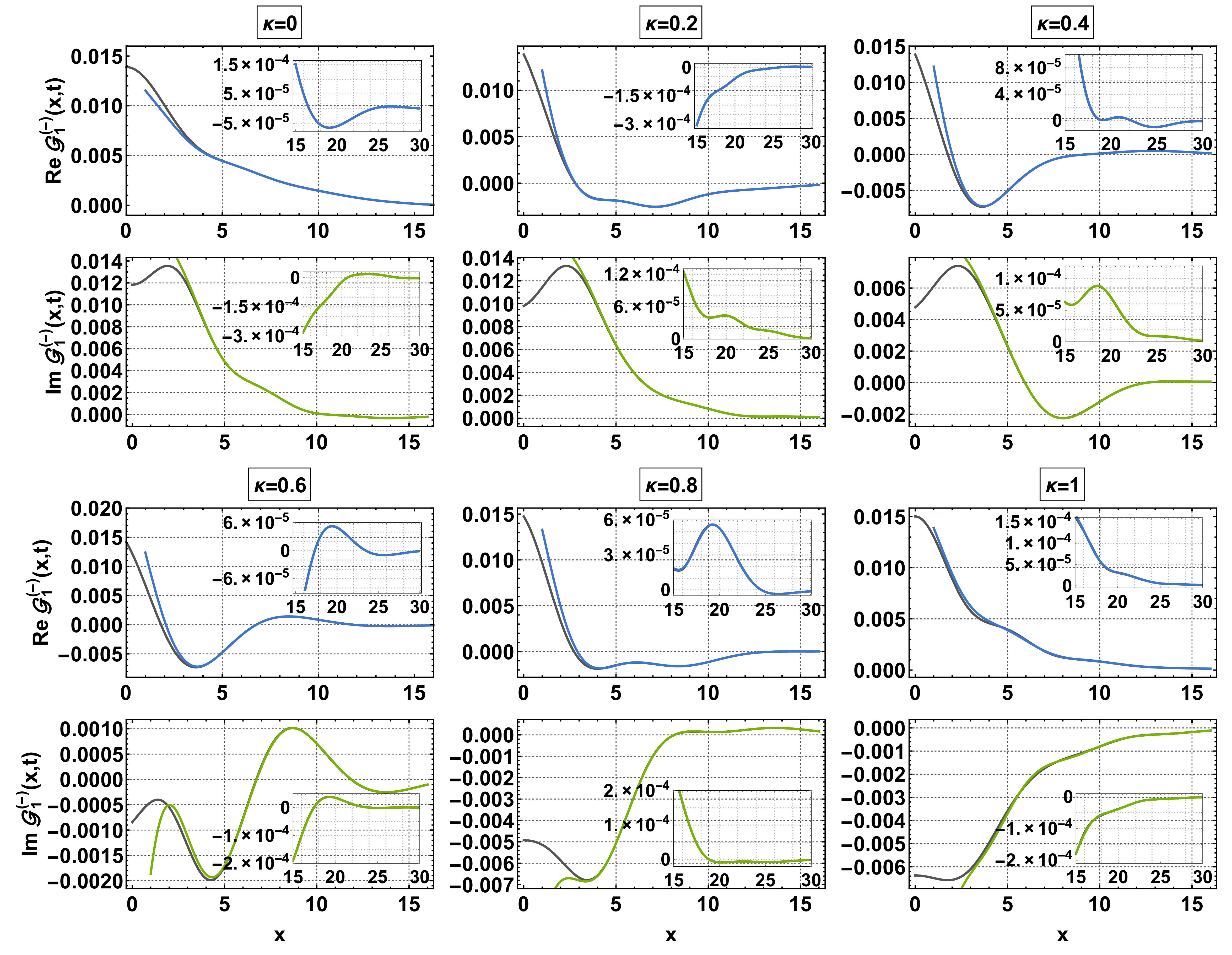}
\caption{Plots of the dynamic correlator $\mathcal{G}_1^{(-)}(x,t)$  at fixed $t=25$ and $h=2$ (black line) and the asymptotic formula
(\ref{gminustime}) (real part blue line, imaginary part green line) for the time-like region. }
\label{gmtime}
\end{figure*}
\begin{figure*}
\includegraphics[width=0.71 \linewidth]{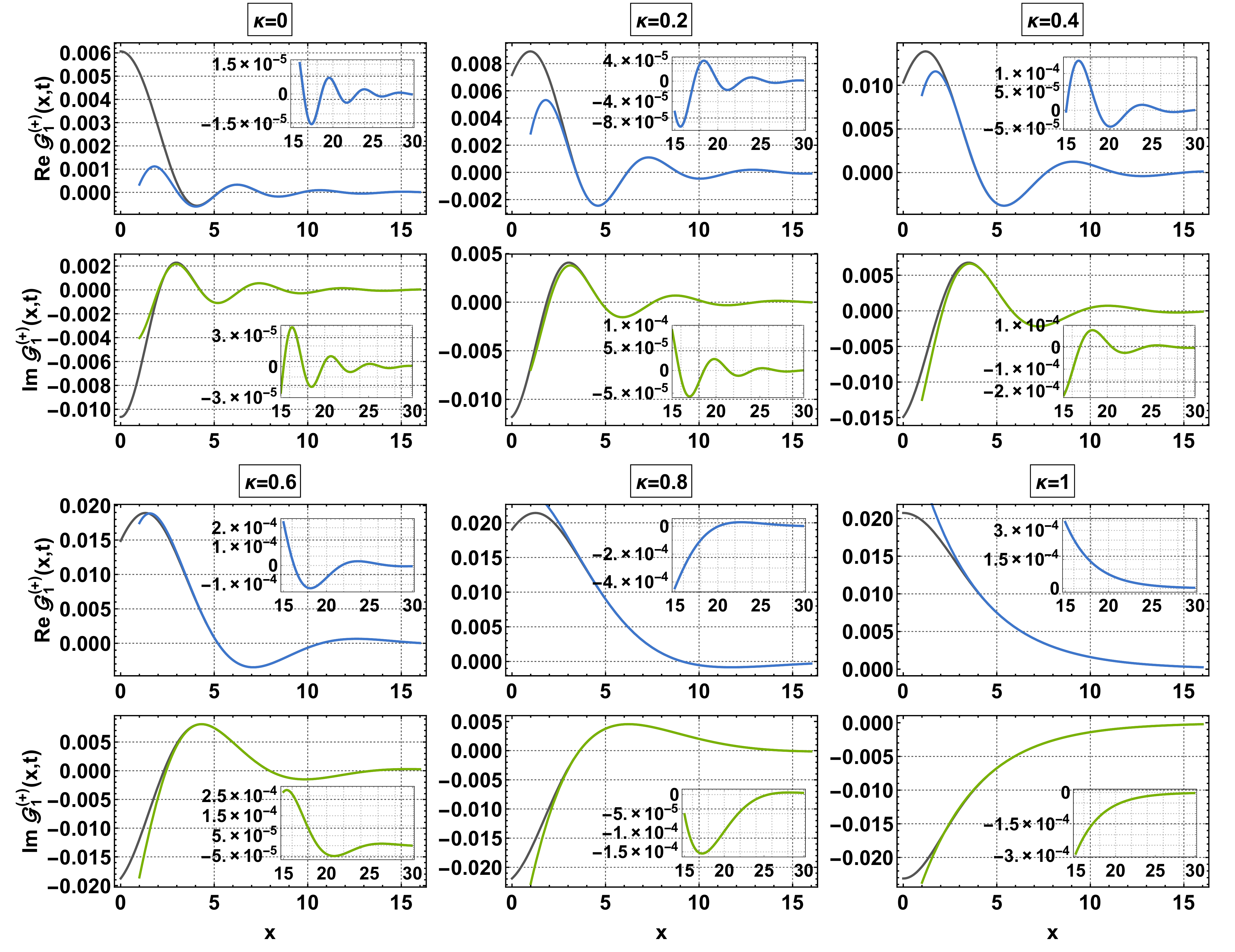}
\caption{Plots of the dynamic correlator $\mathcal{G}_1^{(+)}(x,t)$  at fixed $t=25$ and $h=2$ (black line) and the asymptotic formula
(\ref{gplustime}) (real part blue line, imaginary part green line) for the time-like region. }
\label{gptime}
\end{figure*}

\textit{Results in the time-like regime.} The solution of the RHP in the time-like regime is sketched in Appendix \ref{a4}. For the $\mathcal{G}_1^{(-)}(x,t)$ correlator using the
the representation (\ref{i6}) where the second term is negligible in the large $x$ limit, Thm.~\ref{thmd} and (\ref{Bmmtime}) we find

\begin{flalign}\label{gminustime}
\mathcal{G}^{(-)}_1(x,t)&=
e^{\mathcal{C}(\nu)}e^{2i k_F x\nu}\left\{-(e^{-2i\pi \nu}-1)^2\frac{e^{-i\pi\nu^2}e^{2 i\pi\nu}}{8\sin^2\pi\nu}
\frac{e^{-i th}\overline{G}(x,t)}{(2k_F t+x)^{\nu^2}(2k_F t-x)^{\nu^2}}
\left(\frac{2k_F t+x}{2k_F t-x}\right)^{2\nu}\, \right.
\qquad\qquad\qquad\qquad\qquad\qquad\qquad\qquad\qquad\qquad\qquad\nonumber\\
&+\frac{\pi \left(e^{-2i\pi\nu}-1\right)}{8 e^{i\pi\nu^2}\sin^3\pi\nu}
\left[
\frac{(2k_F)^{-2\nu}}{\Gamma^2(-\nu)}\frac{e^{i k_F x}}{(2k_F t-x)^{\nu^2}(2k_F t+x)^{(\nu+1)^2}}
\left.
+\frac{(2k_F)^{2\nu}}{\Gamma^2(\nu)}\frac{e^{2i\pi\nu}e^{-i k_F x}}{(2k_F t-x)^{(\nu-1)^2}(2k_F t+x)^{\nu^2}}
\right]\right\}\, ,
\end{flalign}
with $\nu=\frac{i\ln 2}{2\pi}+\frac{\kappa}{2}\, .$
In the case of  the $\mathcal{G}^{(+)}_1(x,t)$ correlator using the representation (\ref{i7}), Thm.~\ref{thmd} and (\ref{bpptime}) we obtain
\begin{align}\label{gplustime}
\mathcal{G}^{(+)}_1(x,t)=-e^{\mathcal{C}(\nu)}\pi e^{-i\pi\nu^2}\frac{ \left(e^{-2i\pi\nu}-1\right)^{-1}}{\sin\pi\nu}&e^{2ik_F\nu x}
\left[
\frac{(2k_F)^{2\nu}}{\Gamma^2(\nu)}\frac{e^{-i k_F x}}{(2k_F t-x)^{\nu^2}(2k_F t+x)^{(\nu-1)^2}}\right.\nonumber\\
&\qquad\qquad\qquad\qquad\left.
+\frac{(2k_F)^{-2\nu}}{\Gamma^2(-\nu)}\frac{e^{-2i\pi\nu}e^{i k_F x}}{(2k_F t-x)^{(\nu+1)^2}(2k_F t+x)^{\nu^2}}
\right]\,  ,
\end{align}
with $\nu=\frac{i\ln 2}{2\pi}-\frac{\kappa}{2}$. In the fermionic case similar results (modulo a constant)  were derived in \cite{CZ2}. Our
results contain this previously unknown constant and are also valid for all values of the statistics parameter. Plots of  the corelators for
$h=2$ and fixed $t=25$ together with the expansions (\ref{gminustime}) and (\ref{gplustime}) are presented in   Figs.~\ref{gmtime}  and \ref{gptime}.

\begin{figure*}
\includegraphics[width=1 \linewidth]{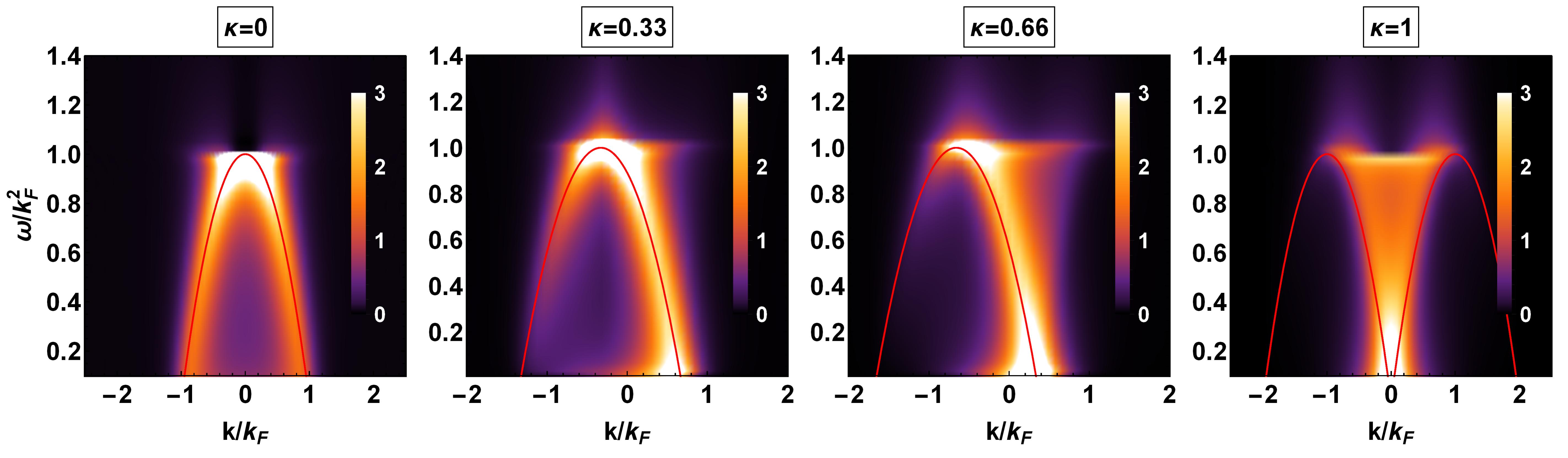}
\caption{
Time and space Fourier transform of $\mathcal{G}_1^{(-)}(x,t)$ at $T=0\, , B=0$ and $h=2$ for several values of the statistics parameter. The red curves are
$\omega(k)=-(k-\kappa k_F)^2 +k_F^2$ for $\kappa=\{0,0.33,0.66\}$ and $\omega(k)=-(k\pm\kappa k_F)^2 +k_F^2$ for $\kappa=1$ (bosons).
 }
\label{FigFourier}
\end{figure*}

The time and space Fourier transform of the correlator defined by
\be
\tilde{\mathcal{G}}_1^{(-)}(\omega,k)=\inti\inti e^{i(\omega t - k x)}\,\mathcal{G}_1^{(-)}(x,t)\, dx\, dt\, ,
\ee
is shown in Fig.~\ref{FigFourier}. The asymmetry in $k$ can be  understood
using the symmetry relations (\ref{gensymm}) and rewriting the Fourier transform as
\begin{align}
\tilde{\mathcal{G}}_1^{(-)}(\omega,k)&=\int_0^\infty\int_0^\infty
2\cos(\omega t - k x)\mbox{Re}\,\mathcal{G}_1^{(-)}(x,t|\kappa)+2\cos(\omega t + k x)\mbox{Re}\,\mathcal{G}_1^{(-)}(x,t|-\kappa)\nonumber\\
&\qquad\qquad\ \ \  -2\sin(\omega t - k x)\mbox{Im}\,\mathcal{G}_1^{(-)}(x,t|\kappa)-2\sin(\omega t + k x)\mbox{Im}\,\mathcal{G}_1^{(-)}(x,t|-\kappa)
\, dx\, dt\, .
\end{align}
For the bosonic an fermionic systems $(\kappa=0,1$) we have $\mathcal{G}_1^{(-)}(x,t|-\kappa)=\mathcal{G}_1^{(-)}(x,t|\kappa)$  and the last relation implies that
$ \tilde{\mathcal{G}}_1^{(-)}(\omega,k)=\tilde{\mathcal{G}}_1^{(-)}(\omega,-k)$ but for an arbitrary value of $\kappa$ we do not have
equality and therefore the Fourier transform is nonsymmetric.

\section{Asymptotics for the single component case}\label{s5b}

The determinant representation for the correlators of the single component system at zero temperature can be easily
derived from (\ref{1minus}) and (\ref{1plus}) using $\lim_{T\rightarrow 0}\vartheta(k)=\theta(k_F^2-k)$  with
$k_F=h^{1/2}$ and $h$ is the chemical potential of the single component system. The derivation of the asymptotics
is very similar with the one presented in the previous sections for the two-component case, therefore, we only  present the final results.
In the case of the static correlators the large distance asymptotic behavior is given by (we drop the subscript)
\be\label{1corrstatic}
\mathcal{G}^{(-)}(x,0)= \frac{ i e^{\mathcal{C}\left(-\frac{\kappa}{2}\right)}}{2\pi} e^{- i \frac{\pi \kappa}{2}}
\left[   (2k_F)^{\kappa}\Gamma^2\left(1-\frac{\kappa}{2}\right)\frac{e^{i k_F(\kappa-1)x}}{x^{1-\kappa +\kappa^2/2}}
-(2k_F)^{-\kappa}\Gamma^2\left(1+\frac{\kappa}{2}\right)\frac{e^{i k_F(\kappa+1)x}}{x^{1+\kappa +\kappa^2/2}}
\right]\, ,
\ee
which confirms the bosonization result of Calabrese and Mintchev (\ref{1CFT}) \cite{CM} and, in addition, supplies
explicit expressions for the amplitudes.
In the bosonic limit $(\kappa=1)$ the first term on the right hand side of (\ref{1corrstatic}) is dominant and using
$G_B(1/2)=A^{-3/2}\pi^{-1/4}e^{1/8}2^{1/24}$ \cite{Barnes} and $G(3/2)=\Gamma(1/2)G_B(1/2)$ with $A=1.2824\ldots$  Glaisher's
constant we find
$
\mathcal{G}^{(-)}(x,0)=k_F \rho_\infty/\pi |k_F x|^{1/2}\, ,\  \rho_\infty=\pi e^{1/2}2^{-1/3}A^{-6}\, ,
$
which reproduces  the leading term derived by Vaidya and Tracy \cite{VT1,VT2} for impenetrable bosons (see also \cite{JMMS,Gang}).
For $\kappa=0$ both terms are important and we obtain $\mathcal{G}^{(-)}(x,0)=\sin(k_F x)/\pi x$ which is in fact the exact
result for spinless fermions.

The asymptotic expansions of the dynamic correlators are almost identical with the ones derived for the two component
case the main difference appearing in the value of the $\nu$ parameter. More precisely, in the single component case the asymptotics
for  $\mathcal{G}^{(-)}(x,t)$  are given by  (\ref{gminusspace}), (\ref{gminustime}) multiplied by 2 with $\nu^{(-)}=\kappa/2$  while for
$\mathcal{G}^{(+)}(x,t)$  the asymptotics are  (\ref{gplusspace}), (\ref{gplustime})  with the parameter $\nu^{(+)}=-\kappa/2$.

\section{Form factors}\label{s6}

In the following sections we  present the derivation of the determinant representations (\ref{detm}) and (\ref{detp}).
We are going to use the same method employed by the authors of \cite{IP} who  derived the similar representations for bosons
and fermions.  Some of the details of the computations are very similar so we will focus mainly on the particular complications
induced by the anyonic statistics. The derivation consists of three main steps. In the initial step we are going to consider a finite
size system and insert a resolution of identity between the two operators appearing in the definition of the correlators (\ref{defcorr}).
Each term in the trace can be written as a sum over the form factors which are the building blocks of our representation. For example,
in the finite system the $\Gm{\beta}(x,t)$ correlators take the form
\be
\Gm{\beta}(x,t)=
\frac{\sum e^{-E_{N+1,M}(\bm{k})/T}\langle\psi^\dagger_\beta(x,t)\psi_\beta(0,0)\rangle_{N+1,M}}{\sum e^{-E_{N+1,M}(\bm{k})/T}}\, ,\ \ \beta \in\{1,2\}\, , \\
\ee
with $\langle\psi^\dagger_\beta(x,t)\psi_\beta(0,0)\rangle_{N+1,M}$ the normalized value of the operators $\psi^\dagger_\beta(x,t)\psi_\beta(0,0)$
in the state $|\Psi_{N+1,M}(\bm{k},\bm{\lambda})\rangle$. Insertion of the identity produces (\ref{i12}) with the form factors defined in (\ref{detff}).
The form factors can be expressed as finite size determinants as it will be shown in this section. The second step is the summation of form factors
using the ``summation under the determinant" trick (see Section \ref{s7}). The final step is taking the thermodynamic limit which will be done in Section \ref{s8}.

\subsection{Determinant representation for the form factors}

In this section we are going to obtain the determinant representation for the form factors in the finite system.
The form factors are defined as
\be\label{deff}
\mathcal{F}_{N,M}^{(\beta)}(x,t)\equiv\langle\Psi_{N,\bar{M}}(\bm{q},\bm{\mu})|\psi_\beta(x,t)|\Psi_{N+1,M}(\bm{k},\bm{\lambda)}\rangle\, ,\ \ \beta=\{1,2\}\, ,
\ee
where $\bar{M}=M$ if $\beta=1$ and $\bar{M}=M-1$ if $\beta=2$. The form factor of the creation operator $\psi_\beta^\dagger(x,t)$
is given by the complex conjugate of $\mathcal{F}_{N,M}^{(\beta)}(x,t)$. In (\ref{deff}) the eigenstate in the $(N+1,M)$-sector is
characterized by the quasimomenta $\bm{k}=\{k_a\}_{a=1}^{N+1}, \bm{\lambda}=\{\lambda_b\}_{b=1}^M$ and the one in the $(N,\bar{M})$-sector by
$\bm{q}=\{q_a\}_{a=1}^N, \bm{\mu_b}=\{\mu_b\}_{b=1}^{\bar{M}}$. They satisfy the
following set of BAEs:
\begin{subequations}\label{baeff}
\begin{align}
 & (N+1,M) \mbox{-sector}  & & &    & (N,\bar{M}) \mbox{-sector} & & &\nonumber\\
e^{i k_a L}         &=\omega e^{-i\pi\kappa N}\, ,&\ \ &a=1,\cdots, N+1\, ,& e^{i q_a L}         &=\zeta e^{-i\pi\kappa (N-1)}\, ,&\ \ &a=1,\cdots, N\, ,&   \\
e^{i\lambda_b (N+1)}&=(-1)^{M-1}\, ,              &\ \ &b=1,\cdots,M\, ,   & e^{i\mu_b N}&=(-1)^{\bar{M}-1}\, ,              &\ \ &b=1,\cdots,\bar{M}\, ,   &
\end{align}
\end{subequations}
where we have introduced
\be\label{defwz}
\omega=e^{i\sum_{b=1}^M \lambda_b}\, ,\ \ \zeta=e^{i\sum_{b=1}^{\bar{M}} \mu_b}\, .
\ee
Using $\psi_\beta^\dagger(x,t)=e^{i t \mathcal{H}} \psi_\beta^\dagger(x) e^{-i t \mathcal{H}}$ and the fact that $|\Psi_{N,\bar{M}}(\bm{q},\bm{\mu})\rangle$
and $|\Psi_{N+1,M}(\bm{k},\bm{\lambda)}\rangle$ are eigenstates of the Hamiltonian with eigenvalues given by (\ref{energy}) we obtain
\be
\mathcal{F}_{N,M}^{(\beta)}(x,t)=\exp\left\{ it\left(\sum_{a=1}^N q_a^2-\sum_{a=1}^{N+1}k_a^2+h_\beta\right)\right\}\mathcal{F}_{N,M}^{(\beta)}(x)\, ,
\ee
where $h_\beta$ is the chemical potential of the  particles $(h_1=h-B, h_2=h+B)$ and we introduced the notation
$\mathcal{F}_{N,M}^{(\beta)}(x,0)=\mathcal{F}_{N,M}^{(\beta)}(x)\, .$ The starting point of our computation is the formula
\footnote{
Note that the wavefunction also satisfies $\chi^{\cdots\alpha_i\alpha_{i+1}\cdots}(\cdots,z_i,z_{i+1},\cdots)=-e^{-i\pi\kappa\, \scriptsize{\mbox{sgn}}(z_i-z_{i+1})}
\chi^{\cdots\alpha_{i+1}\alpha_{i}\cdots}(\cdots,z_{i+1},z_{i},\cdots)$.}
\be
\mathcal{F}_{N,M}^{(\beta)}(x)=(N+1)!\int_0^L dz_1\cdots dz_N\sum_{\alpha_1,\cdots,\alpha_N=\{1,2\}}^{[N,\bar{M}]}\bar{\chi}_{N,\bar{M}}^{\alpha_1\cdots\alpha_N}(\bm{z}|\bm{q},\bm{\mu})
\chi_{N+1,M}^{\alpha_1\cdots\alpha_N\beta}(\bm{z},x|\bm{k},\bm{\lambda})\, ,
\ee
which is obtained by using the commutation relations (\ref{commr}) to move the $\psi_{\beta}(x,0)$ operator to the right in (\ref{deff}) until it hits the vacuum.
In the previous relation $\bm{z}=\{z_1,\cdots,z_N\}$ and the bar denotes complex conjugation. Using the definition of the wavefunctions (\ref{wavef})
we find
\begin{align}\label{i11}
\mathcal{F}_{N,M}^{(\beta)}(x)&=\frac{e^{-i\pi\kappa N}}{N!}\int_0^Ldz_1\cdots dz_N\sum_{R\in S_N}\left\{\theta\left(z_{R(1)}<\cdots<z_{R(N)}<x\right)F_\beta(N)\right.\nonumber\\
&\ \ +\sum_{j=1}^{N-1}\theta\left(z_{R(1)}<\cdots<z_{R(j)}<x<z_{R(j+1)}<\cdots<z_{R(N)}\right)e^{i\pi\kappa(N-j)}F_\beta(j)\nonumber\\
&\ \ \left.+\,\theta\left(x<z_{R(1)}<\cdots<z_{R(N)}\right)e^{i\pi\kappa N} F_\beta(0)\right\}\nonumber\\
&\ \ \times \left(\sum_{Q\in S_N}(-1)^Q e^{-i\left(q_{Q(1)} z_1+\cdots +q_{Q(N)}z_N\right)}\right)\left(\sum_{P\in S_N}(-1)^P e^{i\left(k_{P(1)} z_1+\cdots +k_{P(N)}z_N+k_{P(N+1)} x\right)}\right)\, ,
\end{align}
with $(-1)^Q$ the signature of the permutation and $F_\beta(j)=\sum_{\alpha_1,\cdots\alpha_N}\bar{\xi}_{N,\bar{M}}^{\alpha_1\cdots\alpha_N}
\xi_{N+1,M}^{\alpha_1\cdots\alpha_j\beta\alpha_{j+1}\cdots\alpha_N}.$ Remembering that
$\xi_{N,\bar{M}}^{\alpha_1\cdots\alpha_N}$  and $\xi_{N+1,M}^{\alpha_1\cdots\alpha_{N+1}}$ are eigenvectors of the cyclic shift operators $C_N$ and $C_{N+1}$ acting in
a $2^N$ or $2^{N+1}$ dimensional space we have $\xi_{N,\bar{M}}^{\alpha_1\cdots\alpha_N}=\zeta\,\xi_{N,\bar{M}}^{\alpha_2\cdots\alpha_N\alpha_1}$ and $\xi_{N+1,M}^{\alpha_1\cdots\alpha_{N+1}}=\omega\,\xi_{N+1,M}^{\alpha_2\cdots\alpha_{N+1}\alpha_1}$ with $\omega$ and $\zeta$ defined in (\ref{defwz})
satisfying  $\omega^{N+1}=\xi^N=1$.
Using these relations we obtain $F_\beta(j)=\left(\bar{\omega}\zeta\right)^{N-j}F_\beta(N)$ and the sum over the $R$ permutation can be written as
\begin{align}
\sum_{R\in S_N}&\left\{\theta\left(z_{R(1)}<\cdots<z_{R(N)}<x\right)\right.
 +\sum_{j=1}^{N-1}\theta\left(z_{R(1)}<\cdots<z_{R(j)}<x<z_{R(j+1)}<\cdots<z_{R(N)}\right)\left(\bar{\omega}\zeta e^{i\pi\kappa}\right)^{N-j}\nonumber\\
&\ \ \left.+\,\theta\left(x<z_{R(1)}<\cdots<z_{R(N)}\right)\left(\bar{\omega}\zeta e^{i\pi\kappa}\right)^{N}\right\} F_\beta(N)\, .
\end{align}
Introducing a generalization of the sign function $\rho(z)=\theta(z)+e^{i\pi\kappa}\bar{\omega}\zeta\theta(z)$ the previous sum over permutation is
$\sum_{R\in S_N}\{\cdots\}=\prod_{j=1}^N\rho(x-z_j)\, $ (the value of $\rho(0)$ is not important because when two $z$'s are equal the determinants
in (\ref{i11}) vanish). Therefore, (\ref{i11}) becomes
\be
\mathcal{F}_{N,M}^{(\beta)}(x)=\frac{e^{-i\pi\kappa N}}{N!}F_\beta\int_0^L dz_1\cdots dz_N\prod_{j=1}^N \rho(x-z_j)
\sum_{\substack{P\in S_{N+1}\\Q\in S_N}} (-1)^{P+Q}\left(\prod_{j=1}^N e^{i(k_{P(j)}-q_{Q(j)}) z_j}\right) e^{ik_{P(N+1)}x}\, ,
\ee
where we have introduced the notation $F_\beta=F_\beta(N)$.  The integrand is now factorized so the multiple integral can be calculated using
the  formula (the BAEs (\ref{baeff}) have to be taken into account)
\be
\int_0^L\rho(x-z)e^{i(k-q)z}\, dz=-i(1-e^{i\pi\kappa}\bar{\omega}\zeta)\frac{e^{i(k-q)x}}{k-q}\, .
\ee
We find
\be
\mathcal{F}_{N,M}^{(\beta)}(x)=e^{-i\pi\kappa N}F_\beta(-i)^N(1-e^{i\pi\kappa}\bar{\omega}\zeta)^Ne^{i\left(\sum_{a=1}^{N+1}k_a-\sum_{a=1}^N q_a\right)x}
\sum_{P\in S_{N+1}}(-1)^P\prod_{j=1}^N\frac{1}{k_{P(j)}-q_j}\, ,
\ee
where it only remains to derive the auxiliary lattice form factors $F_\beta$. The derivation is the same as in the  bosonic and fermionic case and can be read
directly from \cite{IP} (see formulae 3.23 and 3.25). Collecting all the results we can state the main result of this section. The form factors of a system of
impenetrable two-component anyons admit the following determinant representation in the finite box
\be\label{detff}
\mathcal{F}_{N,M}^{(\beta)}(x)=(-ie^{-i\pi\kappa})^N(1-e^{i\pi\kappa}\bar{\omega}\zeta)^N \mbox{\textsf{det}}_M B_\beta\, \mbox{\textsf{det}}_{N+1} D \,
\exp\left\{\sum_{a=1}^{N+1}(-it k_a^2+i x k_a)-\sum_{j=a}^N (-i t q_a^2+i x q_a)+i t h_\beta\right\}
\, ,
\ee
with $h_1=h-B,$ $h_2=h+B$ and $D$ is a $(N+1)\times(N+1)$ matrix with elements
\be
D_{ab}=\frac{1}{k_a-q_b}\, ,\ \ a=1,\cdots,N+1\, , \ \ \ \ \  D_{a,N+1}=1\, ,\ \ b=1,\cdots,N\, .
\ee
The $B_1$  and $B_2$  are $M\times M$ matrices with elements
\begin{align}
[B_1]_{ab}&=\sum_{n=1}^N e^{i(\lambda_a-\mu_b) n}\, ,\ \ a,b=1,\cdots,M\, ,\\
[B_2]_{ab}&=\sum_{n=1}^N e^{i(\lambda_a-\mu_b) n}\, ,\ \ [B_2]_{aM}=1\, ,\ a=1,\cdots, M\, ,\ \  b=1, \cdots, M-1\, .
\end{align}
At $\kappa=0$  and $\kappa=1$  (\ref{detff}) reproduces (modulo a phase factor) the results for fermions and bosons derived in \cite{IP}.
For $M=0$ we have $\mathcal{F}_{N,M}^{(2)}=0$, $\omega=\zeta=1$ and $\mbox{\textsf{det}}_M B_1=1$. The form factor $\mathcal{F}_{N,M}^{(1)}$ is in this case
equal again modulo a phase) with the form factor of single component hardcore anyons \cite{PKA3} (note that the statistics parameter used in \cite{PKA3} is related to ours
via $\kappa'=1+\kappa$). If $M=N+1$ then the form factor $\mathcal{F}_{N,M}^{(1)}$ vanishes and $\zeta=\omega=1$. Now the form factor $\mathcal{F}_{N,M}^{(2)}$ reduces to the result for single
component anyons (the $\mbox{\textsf{det}}_{M=N+1} B_2$ factor is related to the normalization of the eigenstates and irelevant).

\section{Summation of form factors}\label{s7}

The second step in our derivation of the determinant representations (\ref{detm}) and (\ref{detp}) is represented by the
calculation of normalized mean values of bilocal operators by summation of the form factors. From now on we will focus on
the correlators of the type 1 particles as the second set of correlators can be obtained using the relation (\ref{symm}).
An independent calculation, which we do not present here, using the form factors for the second type of particles
confirms the symmetry (\ref{symm}) (for the bosonic and fermionic case see \cite{IP}).

\subsection{Summation of form factors for $\langle\psi_1^\dagger(x,t)\psi_1(0,0)\rangle_{N+1,M}$}

Inserting a resolution of the identity between the operators of the normalized mean value $\langle\psi_1^\dagger(x,t)\psi_1(0,0)\rangle$ we find
\begin{align}\label{i12}
\langle\psi_1^\dagger(x,t)\psi_1(0,0)\rangle_{N+1,M}&=\frac{\langle\Psi_{N+1,M}(\bm{k},\bm{\lambda})|\psi_1^\dagger(x,t)\psi_1(0,0)|\Psi_{N+1,M}(\bm{k},\bm{\lambda})\rangle}
{\langle\Psi_{N+1,M}(\bm{k},\bm{\lambda})|\Psi_{N+1,M}(\bm{k},\bm{\lambda})\rangle}\, ,\nonumber\\
&=\sum_{\bm{q},\bm{\mu}}^{[N,M]} \frac{\langle\Psi_{N+1,M}(\bm{k},\bm{\lambda})|\psi_1^\dagger(x,t)|\Psi_{N,M}(\bm{q},\bm{\mu})\rangle}
{\langle\Psi_{N,M}(\bm{q},\bm{\mu})|\Psi_{N,M}(\bm{q},\bm{\mu})\rangle}
\frac{\langle\Psi_{N,M}(\bm{q},\bm{\mu})|\psi_1(0,0)|\Psi_{N+1,M}(\bm{k},\bm{\lambda})\rangle}
{\langle\Psi_{N+1,M}(\bm{k},\bm{\lambda})|\Psi_{N+1,M}(\bm{k},\bm{\lambda})\rangle}\, ,
\nonumber\\
&=\sum_{\bm{q},\bm{\mu}}^{[N,M]} \frac{\bar{\mathcal{F}}_{N,M}^{(1)}(x,t) \mathcal{F}_{N,M}^{(1)}(0,0)}{L^{2N+1}N^M(N+1)^M}\, ,
\end{align}
where we have used the normalization of the eigenstates (\ref{normalization}) and the form factors  defined in (\ref{detff}). The summation
in (\ref{i12}) is over all allowed values of $\bm{q}$ and $\bm{\mu}$ in the $(N,M)$-sector. Introducing
\be
\Lambda=\sum_{a=1}^M\lambda_a\, ,\ \ \ \Theta=\sum_{a=1}^M\mu_a\, ,
\ee
we see from (\ref{allowed}) that the allowed values of the quasimomenta can be written as $q_a=\tilde{q}_a+\Theta/L$ and $k_a=\tilde{k}_a+\Lambda/L$
with $\tilde{q}_a$ not depending on $\mu$'s and $\tilde{k}_a$ on $\lambda$'s. This means that we can sum over $\tilde{q}_a$ independently on $\mu$'s.
The summand in (\ref{i12}) is symmetric in both $q$'s and $\tilde{q}$'s and also on $\mu$'s and it vanishes when two $\mu$'s or $q$'s coincide (because
it involves square moduli of determinants). Therefore, the sum can be written as
\be
\sum_{\bm{q},\bm{\mu}}\equiv\sum_{\substack{q_1<\cdots<q_N\\ \mu_1<\cdots< \mu_M}}\rightarrow\frac{1}{N!}\sum_{\tilde{q}_1}\cdots\sum_{\tilde{q}_N}
\frac{1}{M!}\sum_{\mu_1}\cdots\sum_{\mu_M}
\ee
where each sum is independent of the others. For an arbitrary function $f$ the summation over $\tilde{q}_a$  or $\mu_b$ is given by
\be
\sum_{\tilde{q}_a}f(q_a)=\sum_{j\in\mathbb{Z}}f([\tilde{q}_a]_j)\, ,\ \
\sum_{\mu_b}=\sum_{l=1}^N f([\mu_b]_l)\, ,
\ee
where the allowed values of the quasimomenta are $[\tilde{q}_a]_j=2\pi(j+\delta)/L$ with $j\in\mathbb{Z}$ and $\delta=\{-\pi\kappa(N-1)]\}$ and
$[\mu_b]_l=2\pi l/N$ with $l=0,\cdots, N-1$. The summation over $\tilde{q}_1,\cdots \tilde{q}_N$ is performed along the same lines as in the case of bosons
and fermions \cite{IP}. We find
\be\label{startsumminus}
\langle\psi_1^\dagger(x,t)\psi_1(0,0)\rangle_{N+1,M}=\frac{e^{-it h_1}}{N^M(N+1)^M}\frac{1}{M!}\sum_{\mu_1}\cdots\sum_{\mu_M}\left|\mbox{\textsf{det}}_M B_1\right|^2
\frac{\6}{\6 z}\mbox{\textsf{det}}_{N+1}\left.\left(S^{(-)}+zR^{(-)}\right)\right|_{z=0}\, ,
\ee
with $S^{(-)}$ and $R^{(-)}$ matrices of dimension $(N+1)\times(N+1)$ with elements
\begin{align}
[S^{(-)}]_{ab}&=e_-(k_a)e_-(k_b)\frac{|1-e^{i\pi\kappa}\bar{\omega}\zeta|^2}{L^2}\sum_{\tilde{q}}\frac{e^{-i tq^2+i x q}}{(k_a-q)(k_b-q)}\, , \ \ q=\tilde{q}+\frac{\Theta}{L}\, ,\label{defsminus}\\
[R^{(-)}]_{ab}&=\frac{e_-(k_a)e_-(k_b)}{L}\, ,\ \ \ e_-(k)=e^{\frac{i tk^2-i kx}{2}}\, .\label{defeminus}
\end{align}
In the finite system the elements of the matrices $S^{(-)}$ and $R^{(-)}$ are well defined due to $k_a\ne q_b$ which can be seen from  the BAEs (\ref{baeff}).
In the thermodynamic limit $k$ and $q$ become arbitrary and $\frac{1}{L}\sum_{\tilde{q}}$ is replaced by $\frac{1}{2\pi}\int\, d\tilde{q}$
which means that poles can appear in the integrands. It is therefore necessary to rewrite the elements of the relevant matrices in a way such
that the thermodynamic limit can be taken without problems. The necessary calculations are presented in Appendix \ref{a5} where it is shown
that the matrix $S^{(-)}$ can be written as
\be
S^{(-)}=I+\frac{1-\cos(\Lambda-\Theta-\pi\kappa)}{2}\,V_1^{(-)}-\frac{\sin(\Lambda-\Theta-\pi\kappa)}{2}\,V_2^{(-)}\, ,
\ee
with $V_{1,2}^{(-)}$ matrices of dimension $(N+1)\times(N+1)$ ($I$ is the identity matrix of the same dimensionality) defined by
\begin{align}\label{defvminus}
[V_1^{(-)}]_{ab}&=\frac{2}{L}\frac{e_+^{(-)}(k_a)e_-(k_b)-e_-(k_a)e_+^{(-)}(k_b)}{k_a-k_b}\, ,\ \ \
[V_2^{(-)}]_{ab}=\frac{2}{L}\frac{[e_-(k_a)]^{-1}e_-(k_b)-e_-(k_a)[e_-(k_b)]^{-1}(k_b)}{k_a-k_b}\, .
\end{align}
The function $e_+^{(-)}(k)$ is defined by
\be\label{defeminusupm}
e_+^{(-)}(k)=e^{(-)}(k)e_-(k)\, ,\ \ e^{(-)}(k)=\frac{2}{L}\sum_{\tilde{q}} \frac{e^{-it q^2+iqx}-e^{-it k^2+ikx}}{q-k}\, ,
\ee
with $e_-(k)$ given in (\ref{defeminus}). All the functions appearing in (\ref{defvminus}) and in (\ref{defeminus}) for the $R^{(-)}$
matrix have a well defined thermodynamic limit. Now it only remains to sum over the $\mu$'s in (\ref{startsumminus}). This step is
identical with the one in \cite{IP} and we obtain
\be\label{finalminus}
\langle\psi_1^\dagger(x,t)\psi_1(0,0)\rangle_{N+1,M}=\frac{e^{- i t h_1}}{N}\sum_{n,p=0}^{N-1}e^{-\frac{2 i\pi p}{N} n}
\mbox{\textsf{det}}_M\, \mathcal{U}_p^{(1,-)}\left[\mbox{\textsf{det}}_{N+1}\left(S_n^{(-)}+R^{(-)}\right)-\mbox{\textsf{det}}_{N+1} S_n^{(-)}\right]\, ,
\ee
with $\mathcal{U}_p^{(1,-)}$ a square matrix of dimension $M$  and elements
\be
[\mathcal{U}_p^{(1,-)}]_{ab}=\frac{1}{N(N+1)}\sum_{n=1}^N\sum_{m=1}^N\sum_\mu e^{i (p+m-n)\mu+i n\lambda_a-i m\lambda_b}\, , \ \ a,b=1,\cdots, M\, .
\ee
In (\ref{finalminus}), $S_n^{(-)}=I+\frac{1-\cos(\Lambda-2\pi n/N-\pi\kappa)}{2}\,V_1^{(-)}-\frac{\sin(\Lambda-2\pi n/N-\pi\kappa)}{2}\,V_2^{(-)}$ and also
one should put $q=\tilde{q}+2\pi n/(LN)$ in the function $e^{(-)}(k)$.

\subsection{Summation of form factors for $\langle\psi_1(x,t)\psi_1^\dagger(0,0)\rangle_{N,M}$}

After the insertion of the identity we have
\begin{align}
\langle\psi_1(x,t)\psi_1^\dagger(0,0)\rangle_{N,M}&=\frac{\langle\Psi_{N,M}(\bm{q},\bm{\mu})|\psi_1(x,t)\psi_1^\dagger(0,0)|\Psi_{N,M}(\bm{q},\bm{\mu})\rangle}
{\langle\Psi_{N,M}(\bm{q},\bm{\mu})|\Psi_{N,M}(\bm{q},\bm{\mu})\rangle}\, ,\nonumber\\
&=\sum_{\bm{k},\bm{\mu}}^{[N+1,M]}\frac{\langle\Psi_{N,M}(\bm{q},\bm{\mu})|\psi_1(x,t)|\Psi_{N+1,M}(\bm{k},\bm{\lambda})\rangle}
{\langle\Psi_{N,M}(\bm{q},\bm{\mu})|\Psi_{N,M}(\bm{q},\bm{\mu})\rangle}
\frac{\langle\Psi_{N+1,M}(\bm{k},\bm{\lambda})|\psi_1^\dagger(0,0)|\Psi_{N,M}(\bm{q},\bm{\mu})\rangle}
{\langle\Psi_{N+1,M}(\bm{k},\bm{\lambda})|\Psi_{N+1,M}(\bm{k},\bm{\lambda})\rangle}\, ,\nonumber\\
&=\sum_{\bm{k},\bm{\mu}}^{[N+1,M]}\frac{\bar{\mathcal{F}}_{N,M}^{(1)}(0,0)\mathcal{F}_{N,M}^{(1)}(x,t)}{L^{2N+1}N^M(N+1)^M}\, ,
\end{align}
with the summation over all allowed values of $\bm{k}$ and $\bm{\lambda}$ in the $(N+1,M)$-sector. Again due to the symmetry of the summand
we have $(k=\tilde{k}+\Lambda/L)$
\be
\sum_{\bm{k},\bm{\lambda}}\equiv\sum_{\substack{k_1<\cdots<k_{N+1}\\ \lambda_1<\cdots< \lambda_M}}\rightarrow\frac{1}{N+1!}\,\sum_{\tilde{k}_1}\cdots\sum_{\tilde{k}_{N+1}}
\frac{1}{M!}\,\sum_{\lambda_1}\cdots\sum_{\lambda_M}\, ,
\ee
with each summation independent of the others. Summation over $\tilde{k}_1,\cdots,\tilde{k}_{N+1}$ gives \cite{IP}
\be
\langle\psi_1(x,t)\psi_1^\dagger(0,0)\rangle_{N,M}=\frac{e^{i t h_1}}{N^M(N+1)^M}\frac{1}{M!}\sum_{\lambda_1}\cdots\sum_{\lambda_M}\,|\mbox{\textsf{det}}_M B_1|^2
\left[g(x,t)+\frac{\6}{\6 z}\right]\mbox{\textsf{det}}_N \left.\left(S^{(+)}-z R^{(+)}\right)\right|_{z=0}
\ee
with $g(x,t)=\sum_{\tilde{k}}e^{-i t k^2+i x k}/L$ and $S^{(+)}$ and $R^{(+)}$ are square matrices of dimension $N$ with elements
\begin{align}
[S^{(+)}]_{ab}&=e_-(q_a)e_-(q_b)\frac{|1-e^{i\pi\kappa}\bar{\omega}\zeta|^2}{L^2}\sum_{\tilde{k}}\frac{e^{-i tk^2+i x k}}{(k-q_a)(k-q_b)}\, , \ \ k=\tilde{k}+\frac{\Lambda}{L}\, ,\label{defsplus}\\
[R^{(+)}]_{ab}&=\frac{|1-e^{i\pi\kappa}\bar{\omega}\zeta|^2}{L^3}\left(e_-(q_a)\sum_{\tilde{k}}\frac{e^{-i t k^2+i x k}}{k-q_a}\right)
\left(e_-(q_b)\sum_{\tilde{k}}\frac{e^{-i t k^2+i x k}}{k-q_b}\right)\, .\label{defrplus}
\end{align}
Similar to the previous case the $S^{(+)}$ matrix can be rewritten in a form which is suitable to taking the thermodynamic limit (see Appendix \ref{a5}).
Introducing the function $e_+^{(+)}(q)$ defined by
\be
e_+^{(+)}(k)=e_-(q)e^{(+)}(q)\, ,\ \ \ e^{(+)}(q)=\frac{2}{L}\sum_{\tilde{k}}\frac{e^{-i t k^2+i k x}-e^{-i t q^2+i q x}}{k-q}\, ,
\ee
the $S^{(+)}$ matrix can be written as
\be
S^{(+)}=I+\frac{1-\cos(\Lambda-\Theta-\pi\kappa)}{2}\,V^{(+)}_1+\frac{\sin(\Lambda-\Theta-\pi\kappa)}{2}\,V_2^{(+)}\, ,
\ee
with $V_{1,2}^{(+)}$ square matrices of dimension $N$ and elements
\begin{align}\label{defvplus}
[V^{(+)}_1]_{ab}&=\frac{2}{L}\frac{e_+^{(+)}(q_a)e_-(q_b)-e_-(q_a)e_+^{(+)}(q_b)}{q_a-q_b}\, ,\ \ \
[V^{(+)}_2]_{ab}=\frac{2}{L}\frac{[e_-(q_a)]^{-1}e_-(q_b)-e_-(q_a)[e_-(q_b)]^{-1}}{q_a-q_b}\, .
\end{align}
In the case of the $R^{(+)}$ matrix we obtain
\begin{align}\label{defrplusl}
[R^{(+)}]_{ab}&=\frac{1-\cos(\Lambda-\Theta-\pi\kappa)}{2}\frac{e^{(+)}_+(q_a)e^{(+)}_+(q_b)}{L}
+\frac{\sin(\Lambda-\Theta-\pi\kappa)}{2}\left[\frac{e_+^{(+)}( q_a)}{L e_-(q_b)}+\frac{e_+^{(+)}(q_b)}{L e_-(q_a)}\right]\nonumber\\
&\qquad\qquad\qquad\qquad\qquad +\frac{1+\cos(\Lambda-\Theta-\pi\kappa)}{2}\frac{1}{Le_-(q_a)e_-(q_b)}\, .
\end{align}
All the functions appearing in (\ref{defvplus}) and (\ref{defrplusl}) have a well define thermodynamic limit. The final step is the
summation over $\lambda_1,\cdots\lambda_M$ with the result
\be\label{finalplus}
\langle\psi_1(x,t)\psi_1^\dagger(0,0)\rangle_{N,M}=\frac{e^{i t h_1}}{N+1}\sum_{r,m=0}^N e^{\frac{2\pi i r}{N+1} m}
\mbox{\textsf{det}}_M\, \mathcal{U}_r^{(1,+)}\left[g_m(x,t)+\frac{\6}{\6 z}\right]
\left.\mbox{\textsf{det}}_N\left(S_m^{(+)}-z R_m^{(+)}\right)\right|_{z=0}\, ,
\ee
where the square matrix $\mathcal{U}_r^{(1,+)}$ of dimension $M$  has the elements
\be
[\mathcal{U}_r^{(1,+)}]_{ab}=\frac{1}{N(N+1)}\sum_{m=1}^N\sum_{n=1}^N\sum_\lambda e^{-i(r+m-n)\lambda-i n\mu_a+i m\mu_b}\, , \ \ \ a,b=1,\cdots,M\, .
\ee
In (\ref{finalplus}) the subscript $m$ for $S^{(+)}_m$,  $R^{(+)}_m$ and $g_m(x,t)$ means that in the definitions of these functions $\Lambda$ is
replaced by $2\pi m/(N+1)$.

\section{Thermodynamic limit}\label{s8}

In the thermodynamic limit the allowed values for the quasimomenta fill the entire real axis and the
sums over $\tilde{q}_a$ (or respectively $\tilde{k}_a$) should be replaced by integrals
\be
\frac{1}{L}\sum_{\tilde{q}}f(\tilde{q})\rightarrow\frac{1}{2\pi}\inti f(\tilde{q})\, d\tilde{q}\, .
\ee
The previous relation relies on the fact that for every value of the statistics parameter $[\tilde{q}_a]_{j+1}-[\tilde{q}_a]_{j}=2\pi/L$
and that $[q_a]_j=2\pi j/L+2\pi\delta/L+\Theta/L$ where the last two terms  can be neglected in the thermodynamic limit. The grand canonical
potential which describes the thermodynamics of the system is independent of the statistics and is given by Takahashi's formula
first derived in the context of impenetrable fermions (Gaudin-Yang model) \cite{Tak1,IP}
\be\label{takahashi}
\phi(h,B,T)=-\frac{T}{2\pi}\inti\ln\left(1+2\cosh(B/T)e^{-(q^2-h)/T}\right)\, dq\, .
\ee
The thermodynamic limit of (\ref{finalminus}) and  (\ref{finalplus}) is performed along the same
lines as in the case of the two-component fermions and bosons \cite{IP} taking into account that
$g(x,t)\rightarrow G(x,t)=\inti e^{-it k^2+i xk}/2\pi$ and
\be
e^{(\pm)}(k)\rightarrow\mbox{p.v.}\inti \frac{1}{\pi}\frac{e^{-itq^2+i xq}}{q-k}\, dq\, ,
\ee
obtaining (\ref{detm}) and (\ref{detp}).

\section{Conclusions}\label{s9}

In this paper we have investigated the correlation functions of a 1D two-component system of anyons with strong repulsive contact
interactions. This model is the anyonic generalization of the Gaudin-Yang model and can also be understood as the continuum limit
of the two-component anyonic Hubbard model (2AHM) \cite{CGS1,GCS2}. We have derived determinant representations for the temperature-,
time-, and space-dependent correlation functions which can be straightforwardly implemented numerically. The low energy asymptotics
of the  correlators at zero temperature and zero magnetic field reveal the spin-incoherent nature of the system. The asymptotics present
exponential decay modulated by an oscillatory component with frequency proportional with the statistics parameter and an algebraically
decaying component with anomalous exponents which do not correspond to any unitary conformal field theory. The momentum distribution
and the Fourier transform of the dynamic field-field correlator
are not symmetric in momentum due to the broken space-reversal symmetry. The tails of the momentum distribution exhibit the universal
$1/k^4$ behavior with the amplitude given by Tan's contact which is a monotonic function of the statistics parameter. The natural
extension of our work is the consideration of the equivalent lattice system, the 2AHM \cite{CGS1,GCS2}, in the strong repulsive limit.
In the fermionic case, the well known Hubbard model, the determinant representation of the correlators is already known \cite{IPA1,AIP2}
and the long distance asymptotics of the static correlators has also been determined \cite{CZ3}. This will be addressed in  a future publication.

\acknowledgments
Financial support  from the LAPLAS 6 program of the Romanian National Authority for Scientific Research (CNCS-UEFISCDI) is gratefully acknowledged.

\appendix

\section{Fredholm determinants}\label{a1}

Here we present some minimal information on  Fredholm determinants (Chap. III of \cite{CH}). Consider a
Fredholm integral equation of the second kind
\be
f(x)-\lambda\int_a^b \textsf{K}(x,y) f(y)\, dy=g(x)\, ,
\ee
with $f(x)\, , g(x)$  continuous functions on $[a,b]$ and the kernel $\textsf{K}(x,y)$ continuous, symmetric and bounded.
Introducing the $n$-th iterated kernel
\be
\textsf{K}^{(n)}(x,y)=\int_a^b \textsf{K}(x,z)\, \textsf{K}^{(n-1)}(z,y)\, dz\, ,\ \ \ \textsf{K}^{(1)}(x,y)=\textsf{K}(x,y)\, ,
\ee
and the trace of the operator and its powers
$\mbox{Tr}\, \textsf{K}=\int_a^b \textsf{K}(x,x)dx\, , $ $\mbox{Tr}\,
\textsf{K}^{2}=\int_a^b \int_a^b \textsf{K}(x,y)\,\textsf{K}(y,x)\, dx\,dy\, $ and so on,
the Fredholm determinant of the integral operator $(1-\lambda\,\hat{\textsf{K}})$ is given by
\be
\mbox{\textsf{det}}\left(1-\lambda\, \hat{\textsf{K}}\right)=\sum_{n=1}^\infty (-1)^n\frac{\lambda^n}{n!}\int_a^b dx_1 \cdots\int_a^b dx_n\,
 \textsf{K}_n\, \left(\begin{array} {l}
                             x_1,\ldots, x_n\\
                             y_1,\ldots, y_n\\
                  \end{array} \right)\, ,
\ee
where
\be
 \textsf{K}_n\, \left(\begin{array} {l}
                             x_1,\ldots, x_n\\
                             y_1,\ldots, y_n\\
                  \end{array} \right)\equiv \underset{i\le j,k\le n}{\mbox{\textsf{det}}} \textsf{K}(x_j,y_k)\, .
\ee
Two useful formulae are
\be\label{aa1}
\ln \mbox{\textsf{det}}\left(1-\lambda\, \hat{\textsf{K}}\right)=-\sum_{n=1}^\infty\frac{\lambda^n}{n}\, \mbox{Tr}\,  \textsf{K}^n\, ,\ \
\mbox{ and }  \left(1-\lambda\, \hat{\textsf{K}}\right)^{-1}=1+\lambda\, \textsf{K}^{(1)}+\lambda^2\, \textsf{K}^{(2)} +\cdots\, .
\ee

\section{Numerical implementation of Fredholm determinants}\label{a1b}

Fredholm determinants can be numerically evaluated with relative ease using a method which is based on the
classical Nystr\"om method  for the solutions of Fredholm integral equations of the second kind \cite{Born}. For all
the determinants considered in this paper the diagonal elements need to be computed using l'H\^opital rule. In the case
of the dynamical correlators the functions $E^{(\pm)}(x,t\, |k)$  (\ref{defE}) contain principal value integrals
whose numerical evaluation is time consuming. An efficient numerical implementation of these functions is based on the
fact that the Hilbert transform of a gaussian can be expressed in terms of the $\mbox{erf}(x)$ function defined by
$\mbox{erf}(x)=\frac{2}{\sqrt{\pi}}\int_{0}^x e^{- t^2}\, dt\, .$ More precisely, we have \cite{GiamZ}
\[
E(x,t\, |k)\equiv\frac{1}{\pi}\inti \frac{ e^{-i q^2 t+ i  q x}}{q-k}\, dq=i\,   \mbox{erf}\left(\zeta (x-2 k t)\right) e^{-i k^2 t+ i k x}\, ,\ \ t\ne 0\, ,
\]
where $\zeta=e^{- i \frac{\pi}{4}\scriptsize{\mbox{sgn}}(t)}/(2 \sqrt{|t|})$. Making use of $d\, \mbox{erf}(x)/d x=2 e^{-x^2}/\sqrt{\pi}$, the derivative of the previous
function which appears in the diagonal terms of the determinant can be expressed as
\[
\frac{ d E(x,t\, |k)}{d k}=i\, e^{-i k^2 t+ i k x}\left[- \frac{4 i \zeta t }{\sqrt{\pi}}e^{-\zeta^2 (x-2 k t)^2}-(x-2 k t)\,  \mbox{erf}\,(\zeta (x-2 k t)) \right]\, , \ \ t\ne 0\, .
\]
Another useful formula whose proof is similar with  the computation of the Fresnel integrals with the argument tending to infinity is
\[
G(x,t)=\frac{1}{2\pi} \inti  e^{-i k^2 t+ i  k x}\, dk= \frac{e^{- i \frac{\pi}{4}\scriptsize{\mbox{sgn}}(t)}}{2 \sqrt{|t|}} e^{ \frac{i x^2}{4 t}}\, , \ \ t\ne 0\, .
\]

\section{Asymptotic solution of the RHP in the static case}\label{a2}

\begin{figure*}
\includegraphics[width=0.55 \linewidth]{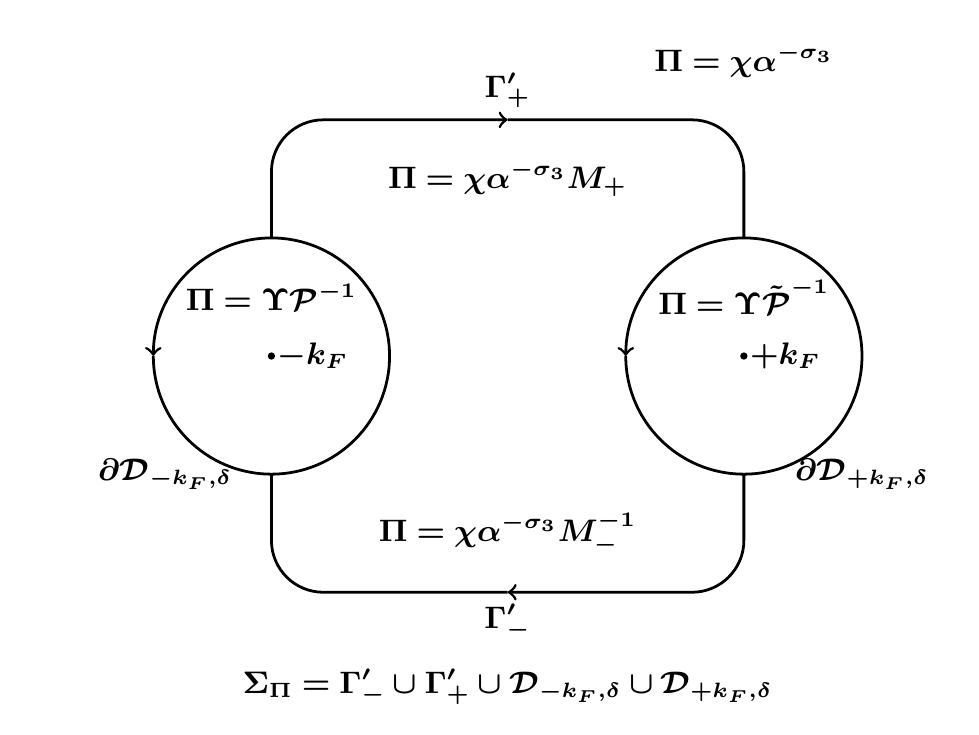}
\caption{Contour for the RHP of $\Pi$ in the static case. The exact definitions of all matrices can be found in \cite{KKMST}. }
\label{frhps}
\end{figure*}

In order to obtain an approximate solution of the RHP for $\chi(k)$ in the large $x$ limit
several transformations are required which are detailed in \cite{KKMST}. In the final step the following
RHP for a matrix  denoted by $\Pi(k)$  is considered (Sect. 4.5 of \cite{KKMST}):

\begin{itemize}
\item $\Pi(k)$ is analytic in $\mathbb{C}\backslash \Sigma_{\Pi}$ where the contour $\Sigma_\Pi=\Gamma_-'\cup\Gamma_+'\cup
\mathcal{D}_{-k_F,\delta}\cup \mathcal{D}_{+k_F,\delta}$ is presented in Fig.~\ref{frhps} and consists of two small circles around $\pm k_F$
of radii $\delta$ and $\Gamma'_\pm$ situated in the upper, respectively, lower half plane.

\item $\Pi(k)\underset{k\rightarrow\infty}{\longrightarrow}I_2+O\left(\frac{1}{k}\right)$ uniformly in $k$.
\item The jump conditions satisfied by $\Pi(k)$ are:

\[
\left\{ \begin{array}{lll}  \Pi_+(k)M_+(k)=\Pi_-(k)       & \mbox{for}   & k\in\Gamma_+'\, ,\\
                            \Pi_+(k)M_-^{-1}(k)=\Pi_-(k)  & \mbox{for}   & k\in\Gamma_-'\, ,\\
                            \Pi_+(k)\mathcal{P}(k)=\Pi_-(k) &\mbox{for}  & k\in\6 D_{-k_F,\delta}\, ,\\
                            \Pi_+(k)\tilde{\mathcal{P}}(k)=\Pi_-(k) &\mbox{for}  & k\in\6 D_{+k_F,\delta}\, .\\
   \end{array} \right.
\]
\end{itemize}

The exact form of the jump matrices $M_{\pm}(k)$, $\mathcal{P}(k),$ and $\tilde{\mathcal{P}}(k)$ is not important for
our analysis (we will however present below the first term in the large $x$ expansion of $\mathcal{P}(k), \tilde{\mathcal{P}}(k)$)
but we should point out that $M_{\pm}(k)$ are exponentially close to $I_2$ on $\Gamma_{\pm}'$ and $\mathcal{P}(k)$ and $\tilde{\mathcal{P}}(k)$
are uniformly $I_2+O\left(x^{\bar{\epsilon}-1}\right)$ on $\6 D_{\pm k_F,\delta}$ with $\bar{\epsilon}= 2\, \mbox{sup}_{\6 D_{\pm k_F,\delta}}|\mbox{Re}(\nu)|$
where $\nu=-\frac{i\ln 2}{2\pi}-\frac{\kappa}{2}$. Outside the contour $\Sigma_\Pi$ we have
\be\label{aa2}
\Pi(k)=\chi(k)\alpha(k)^{-\sigma_3}\, ,\ \ \alpha(k)=\left(\frac{k-k_F}{k+k_F}\right)^\nu\, , \ \ \ \sigma_3=\left(\begin{array}{lr}
                    1&0 \\
                    0&-1
     \end{array}\right)\, .
\ee
Introducing the notation
\be\label{aa3}
\Pi(k)=I_2+\frac{1}{k}
\left(\begin{array}{lr}
                    \Pi_{11}^{(1)}& \Pi_{12}^{(1)} \\
                    \Pi_{21}^{(1)}& \Pi_{22}^{(1)} \\
                 \end{array}   \right)+O\left(\frac{1}{k^2}\right)\, ,\ \ \ \ \
\chi(k)=I_2+\frac{1}{k}
\left(\begin{array}{lr}
                    \chi_{11}^{(1)}& \chi_{12}^{(1)} \\
                    \chi_{21}^{(1)}& \chi_{22}^{(1)} \\
                 \end{array}   \right)+O\left(\frac{1}{k^2}\right)\, ,
\ee
expanding (\ref{aa2}) in powers of $1/k$ and equating terms of the same order we obtain
\[
\left(\begin{array}{lr}
                    \chi_{11}^{(1)}& \chi_{12}^{(1)} \\
                    \chi_{21}^{(1)}& \chi_{22}^{(1)} \\
\end{array}   \right)
                 =
\left(\begin{array}{lr}
                    \Pi_{11}^{(1)}-2k_F \nu& \Pi_{12}^{(1)} \\
                    \Pi_{21}^{(1)}& \Pi_{22}^{(1)}+2k_F\nu \\
\end{array}   \right)\, ,
\]
which together with (\ref{i4}) shows that $B_{--}=-2\pi i\, \Pi_{21}^{(1)}/\xi$. Therefore, now we only need to obtain the
first terms of the expansion of $\Pi(k)$ in $k$ and in $x$. The following result can be proved in a similar fashion with
Prop. 6.2 of \cite{K2}

\begin{prop}
The matrix $\Pi(k)$ admits the asymptotic expansion
\be
\Pi(k)=I_2+\sum_{n=1}^M\frac{\Pi^{(n,x)}(k)}{x^n}+o\left(x^{-M-1-\bar{\epsilon}}\right)\, ,
\ee
which is valid uniformly away from $\Sigma_\Pi$. For $k$ belonging to any connected component of $\infty$ in $\mathbb{C}\backslash\Sigma_\Pi$ the
first term of the expansion is given by
\footnote{In Prop. 6.2 of \cite{K2} this term has a minus sign. This is due to the fact that the contours on $\6 D_{\pm k_F,\delta}$
have a different orientation compared with the ones used in \cite{K2}.}
\be\label{aa4}
\Pi^{(1,x)}(k)=\sum_{\epsilon=\pm}\frac{V^{(\epsilon,1)}(\epsilon k_F)}{k-\epsilon k_F}\, ,
\ee
with $((\nu)_n=\Gamma(\nu+n)/\Gamma(\nu))$
\be
V^{(-,1)}(k)=\left(\begin{array}{lr}
                    -i (-\nu)_1^2& b_{12}(k) \\
                   b_{21}(k)& i(\nu)_1^2 \\
\end{array}   \right)\, ,\ \ \
V^{(+,1)}(k)=\left(\begin{array}{lr}
                    -i (\nu)_1^2& \tilde{b}_{12}(k) \\
                   \tilde{b}_{21}(k)& i(-\nu)_1^2 \\
\end{array}   \right)\, ,
\ee
\begin{align}
b_{12}(k)&=-i\, \Gamma^2(1+\nu)\, \frac{\sin \pi\nu}{\pi}\, \frac{e^{- ik_F x}}{\left[x(k_F-k)\right]^{2\nu}}\, ,\ \ \
b_{21}(k)=-i\, \Gamma^{-2}(\nu)\, \frac{\pi}{\sin \pi \nu}  \frac{e^{+ ik_F x}}{\left[x(k_F-k)\right]^{-2\nu}}\, ,\\
\tilde{b}_{12}(k)&=i\, \Gamma^2(1-\nu)\, \frac{\sin \pi\nu}{\pi}\, \frac{e^{+ ik_F x}}{\left[x(k_F+k)\right]^{-2\nu}}\, ,\ \ \
\tilde{b}_{21}(k)=-i\, \Gamma^{-2}(-\nu)\, \frac{\pi}{\sin \pi \nu}  \frac{e^{- ik_F x}}{\left[x(k_F+k)\right]^{2\nu}}\, .
\end{align}

\end{prop}
From (\ref{aa3}) and (\ref{aa4}) we find $\Pi_{21}^{(1)}=\left[V^{(-,1)}_{21}(-k_F)+V^{(+,1)}_{21}(k_F)\right]/x+o(x^{-2-\bar{\epsilon}})$
and, therefore,
\be\label{aa5}
\frac{1}{4\pi} B_{--}= \frac{\pi}{2x \xi \sin\pi\nu}\left(-\frac{1}{\Gamma^2(\nu)}(2k_F x)^{2\nu} e^{i k_F x}
+\frac{1}{\Gamma^2(-\nu)}(2k_F x)^{-2\nu} e^{-i k_F x}
\right)\, .
\ee

\section{Asymptotic solution of the RHP in the dynamic case. The space-like regime}\label{a3}

\begin{figure*}
\includegraphics[width=0.55 \linewidth]{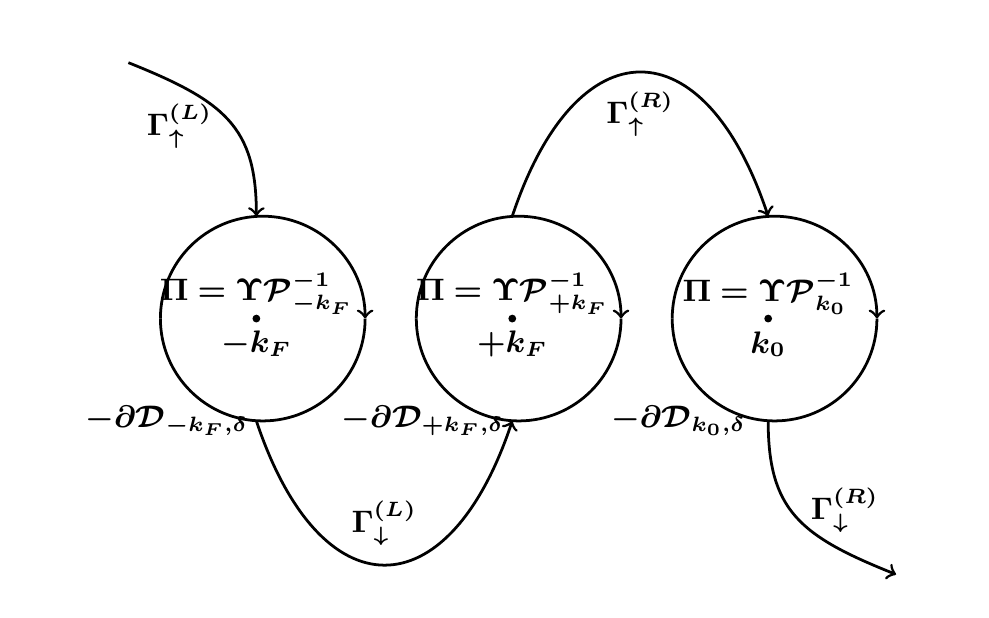}
\caption{Contour $\Sigma_\Pi=\Gamma_\uparrow^{(L)}\cup \Gamma_\downarrow^{(L)}\cup \Gamma_\uparrow^{(R)}  \cup \Gamma_\downarrow^{(R)}\cup
 \mathcal{D}_{-k_F,\delta}\cup \mathcal{D}_{+k_F,\delta}\cup \mathcal{D}_{k_0,\delta}$ for the RHP in the space-like regime.
The exact definitions of all matrices can be found in \cite{K2}.}
\label{frhpdspace}
\end{figure*}

In the space-like regime after a series of transformations Kozlowski \cite{K2} obtained the following RHP:

\begin{itemize}
\item $\Pi(k)$ is analytic in $\mathbb{C}\backslash \Sigma_{\Pi}$ where the contour $\Sigma_\Pi=\Gamma_\uparrow^{(L)}
\cup \Gamma_\downarrow^{(L)}\cup \Gamma_\uparrow^{(R)}  \cup \Gamma_\downarrow^{(R)}\cup \mathcal{D}_{-k_F,\delta}\cup
\mathcal{D}_{+k_F,\delta}\cup \mathcal{D}_{k_0,\delta}$ is presented in Fig.~\ref{frhpdspace}.

\item $\Pi(k)\underset{k\rightarrow\infty}{\longrightarrow}I_2+O\left(\frac{1}{k}\right)$ uniformly in $k$.
\item $\Pi_+(k)G_{\Pi}(k)=\Pi_-(k)$ for $k\in\Sigma_\Pi$ with

\[
\left\{ \begin{array}{lll}  G_{\Pi}(k)=G_\Upsilon(k)      & \mbox{for}   & k\in\Gamma\equiv\Gamma_\uparrow^{(L)}
                                                  \cup \Gamma_\downarrow^{(L)}\cup \Gamma_\uparrow^{(R)}  \cup \Gamma_\downarrow^{(R)}\, ,\\
                            G_{\Pi}(k)=\mathcal{P}_{\pm k_F}^{-1}(k) &\mbox{on}  & k\in-\6 D_{\pm k_F,\delta}\, ,\\
                            G_{\Pi}(k)=\mathcal{P}_{0 }^{-1}(k) &\mbox{on}  & k\in-\6 D_{ k_0,\delta}\, .\\
   \end{array} \right.
\]
\end{itemize}
Again the exact form of the jump matrices is not important for our analysis only the fact that $G_\Upsilon(k)$ is exponentially
close to the unit matrix on $\Gamma$ for large $x$ and that $\mathcal{P}_{\pm k_F}^{-1}(k)$ and $\mathcal{P}_{0 }^{-1}(k)$ present
corrections in powers of $x$ in the same limit on $-\6 D_{\pm k_F,\delta}$ and  $-\6 D_{ k_0,\delta}$. Outside the contour the
connection between $\chi(k)$ and $\Pi(k)$ is given by
\be
\chi(k)=\Pi(k)\alpha^{-\sigma_3}(k)\left(I_2+\sigma^+\mbox{p.v.}\inti\frac{e^{-2}(k)}{k'-k}\frac{dk'}{2\pi i}\right)\, ,\ \
\alpha(k)=\left(\frac{k+k_F}{k-k_F}\right)^{\nu^{(\pm)}}\, , \ \ \ \sigma^+=\left(\begin{array}{lr}
                    0&1 \\
                    0&0
                    \end{array}\right)\, .
\ee
Using the same notations as in (\ref{aa3}), expanding in powers of $1/k$ and equating terms of the same order we obtain
\[
\left(\begin{array}{lr}
                    \chi_{11}^{(1)}& \chi_{12}^{(1)} \\
                    \chi_{21}^{(1)}& \chi_{22}^{(1)} \\
\end{array}   \right)
                 =
\left(\begin{array}{lr}
                    \Pi_{11}^{(1)}+2k_F \nu^{(\pm)}& \Pi_{12}^{(1)}+\frac{1}{2\pi i}\inti e^{-2}(k')\, dk' \\
                    \Pi_{21}^{(1)}& \Pi_{22}^{(1)}-2k_F \nu^{(\pm)} \\
\end{array}   \right)\, .
\]
Together with (\ref{i9}) the potentials $B_{--}$ and $b_{++}$ are expressed in terms of elements of $\Pi^{(1)}$ as
\be\label{aa6}
B_{--}=-\frac{i\pi}{2\sin^2\pi\nu^{(\pm)}}\Pi^{(1)}_{21}\, ,\ \ \ b_{++}=i\Pi^{(1)}_{21}\, .
\ee
The first terms of the large $x$ expansion of $\Pi$ are given by the following proposition:
\begin{prop}\label{prop2}(Prop. 6.2. of \cite{K2}) The matrix $\Pi$ admits the asymptotic expansion
\be
\Pi(k)=I_2+\sum_{n\ge 0}^{N} \frac{\Pi^{(n,x)}}{x^{\frac{1+n}{2}}}
\ee
with the first terms given by
\be
\Pi^{(0,x)}(k)=-\frac{d^{(0)}(k_0)}{k-k_0}\sigma\, ,\ \ \Pi^{(1,x)}(k) =-\sum_{\epsilon=\pm}\frac{V^{(\epsilon,0)}(\epsilon k_F)}{k-\epsilon k_F}\, ,
\ee
when $k$ belongs to any connected component of $\infty$ in $\mathbb{C}\backslash\Sigma_\Pi$. The expansion is valid uniformly away from the contour and $\sigma=\sigma^+$
in the space-like regime and $\sigma=\sigma^-$ in the time like regime.
\end{prop}
In the space-like regime we have $(u(k)=k-t k^2/x)$
\begin{subequations}\label{aa7}
\begin{align}
d^{(0)}(k)&=\alpha^{-2}(k)e^{i x u(k_0)}\frac{\Gamma(1/2)}{2\pi}\frac{e^{i\pi/4}}{t^{1/2}} x^{1/2}\, ,\\
V^{(-,0)}(k)&=-i\left(\frac{k+k_F}{u(k)-u(-k_F)}\right)\left(\begin{array}{cc}
                                                           -(-\nu^{(\pm)})_1^2&- i b_{12}(k)\\
                                                           -i b_{21}(k)      & (\nu^{(\pm)})_1^2
                                                            \end{array}
                                                          \right)\, , \\
V^{(+,0)}(k)&=-i\left(\frac{k-k_F}{u(k)-u(k_F)}\right)\left(\begin{array}{cc}
                                                           -(\nu^{(\pm)})_1^2&- i \tilde{b}_{12}(k)\\
                                                           -i \tilde{b}_{21}(k)      & (-\nu^{(\pm)})_1^2
                                                            \end{array}
                                                          \right)\, ,
\end{align}
\end{subequations}
with
\begin{subequations}\label{aa8}
\begin{align}
b_{12}(k)&=-i \Gamma^2\left(1-\nu^{(\pm)}\right)\frac{\sin\pi\nu^{(\pm)}}{\pi C^{(L)}(k)}\, ,
\ \ b_{21}(k)= -i \frac{\pi C^{(L)}(k)}{\sin\pi\nu^{(\pm)}\Gamma^2\left(-\nu^{(\pm)}\right)}\, ,\\
\tilde{b}_{12}(k)&=+i \Gamma^2\left(1+\nu^{(\pm)}\right)\frac{\sin\pi\nu^{(\pm)}}{\pi C^{(R)}(k)}\, ,
\ \   \tilde{b}_{21}(k)= +i \frac{\pi C^{(R)}(k)}{\sin\pi\nu^{(\pm)}\Gamma^2\left(\nu^{(\pm)}\right)}\, ,
\end{align}
\end{subequations}
and
\begin{subequations}\label{aa9}
\begin{align}
C^{(L)}(k)&=-\left(e^{-2\pi i \nu^{(\pm)}}-1\right)\left(\frac{k+k_F}{u(k)-u(-k_F)}\right)^{2\nu^{(\pm)}}\frac{e^{-i x u(-k_F)}}{\left[x(k_F-k)\right]^{2\nu^{(\pm)}}}\, ,\\
C^{(R)}(k)&=-\left(e^{-2\pi i \nu^{(\pm)}}-1\right)\left(\frac{u(k)-u(k_F)}{k-k_F}+i0^+\right)^{2\nu^{(\pm)}}\frac{e^{-i x u(k_F)}}{\left[x(k_F+k)\right]^{-2\nu^{(\pm)}}}\, .
\end{align}
\end{subequations}
Using (\ref{aa6}), Prop. \ref{prop2} and (\ref{aa7}), (\ref{aa8}) and (\ref{aa9}) we have
\begin{align}\label{Bmmspace}
B_{--}&=\frac{i\pi}{2\sin^2\pi\nu}\frac{1}{x}\left(V^{(-,0)}_{21}(-k_F)+V^{(+,0)}_{21}(k_F)\right)\, ,\ \ \  \nu=\frac{\kappa}{2}-\frac{\eta}{2\pi}\, ,\nonumber\\
&=e^{i t k_F^2}\frac{\pi^2 \left(e^{-2\pi i \nu}-1\right)}{2\sin^2\pi \nu}
\left(\frac{(2k_F)^{-2\nu}}{\sin\pi\nu\, \Gamma^2(-\nu)}\frac{e^{i k_F x}}{(x+2k_F t)^{2\nu+1}}
- \frac{(2k_F)^{2\nu}}{\sin\pi\nu\, \Gamma^2(\nu)}\frac{e^{-i k_F x}}{(x-2k_F t)^{-2\nu+1}}\right)\, .
\end{align}
In a similar fashion
\begin{align}\label{bppspace}
b_{++}&=-\frac{d^{(0)}}{x^{1/2}}-\frac{i}{x}\left(V^{(-,0)}_{12}(-k_F)+V^{(+,0)}_{12}(k_F)\right)\, ,\ \ \  \nu=-\frac{\kappa}{2}+\frac{\eta}{2\pi}\, ,\nonumber\\
&=G(x,t)\left(\frac{x-2k_F t}{x+2k_F t}\right)^{2\nu}-
e^{-i t k_F^2}\frac{\pi \left(e^{-2\pi i \nu}-1\right)^{-1} }{\sin\pi \nu}
\left(
\frac{(2k_F)^{-2\nu}}{\Gamma^2(-\nu)}\frac{e^{i k_F x}}{(x-2k_F t)^{2\nu+1}}
-\frac{(2k_F)^{2\nu}}{ \Gamma^2(\nu)}\frac{e^{-i k_F x}}{(x+2k_F t)^{-2\nu+1}}
\right)\, ,
\end{align}
where we have used that $G(x,t)=e^{ i x^2/(4t)}e^{-i\pi/4}/(2\sqrt{\pi t})$ for $x,t>0$.

\section{Asymptotic solution of the RHP in the dynamic case. The time-like regime}\label{a4}

\begin{figure*}
\includegraphics[width=0.55 \linewidth]{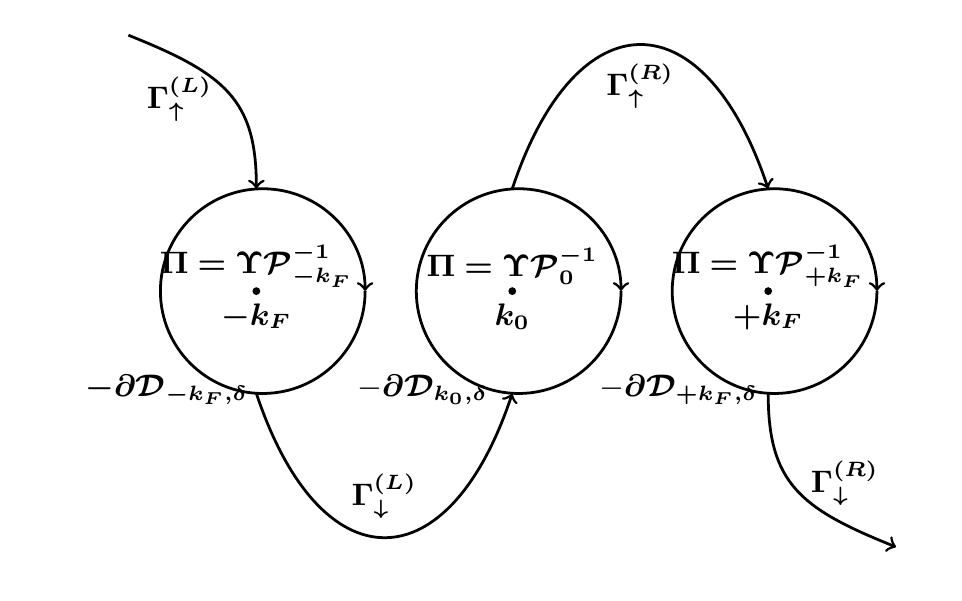}
\caption{Contour $\Sigma_\Pi=\Gamma_\uparrow^{(L)}\cup \Gamma_\downarrow^{(L)}\cup \Gamma_\uparrow^{(R)}  \cup \Gamma_\downarrow^{(R)}\cup
 \mathcal{D}_{-k_F,\delta}\cup \mathcal{D}_{+k_F,\delta}\cup \mathcal{D}_{k_0,\delta}$ for the RHP in the time-like regime.
 The exact definitions of all matrices can be found in \cite{K2}.}
\label{frhpdtime}
\end{figure*}

The RHP for the $\Pi(k)$ matrix in the time-like case is almost the same as the one presented in the previous appendix for the space-like
regime. The contour is presented in Fig.~\ref{frhpdtime} and the only changes are at the level of the parametrices $\mathcal{P}_{\pm k_F}(k)$
and $\mathcal{P}_{ k_0}(k)$.  Prop. \ref{prop2} and (\ref{aa6}) are also true in the time-like regime but (\ref{aa7}) and  (\ref{aa8})
are replaced by
\begin{subequations}\label{aa10}
\begin{align}
d^{(0)}(k)&=-\left(e^{-2i\pi\nu^{(\pm)}}-1\right)^2 \alpha^{2}(k)e^{-i x u(k_0)}\frac{\Gamma(1/2)}{2\pi}\frac{e^{i\pi/4}}{t^{1/2}} x^{1/2}\, ,\\
V^{(-,0)}(k)&=-i\left(\frac{k+k_F}{u(k)-u(-k_F)}\right)\left(\begin{array}{cc}
                                                           -(-\nu^{(\pm)})_1^2&- i b_{12}(k)\\
                                                           -i b_{21}(k)      & (\nu^{(\pm)})_1^2
                                                            \end{array}
                                                          \right)\, , \\
V^{(+,0)}(k)&=-i\left(\frac{k-k_F}{u(k)-u(k_F)}\right)\left(\begin{array}{cc}
                                                           -(\nu^{(\pm)})_1^2&- i \tilde{b}_{12}(k)\\
                                                           -i \tilde{b}_{21}(k)      & (-\nu^{(\pm)})_1^2
                                                            \end{array}
                                                          \right)\, ,
\end{align}
\end{subequations}
where
\begin{subequations}\label{aa11}
\begin{align}
b_{12}(k)&=-i \Gamma^2\left(1-\nu^{(\pm)}\right)\frac{\sin\pi\nu^{(\pm)}}{\pi C^{(L)}(k)}\, ,
\ \ b_{21}(k)= -i \frac{\pi C^{(L)}(k)}{\sin\pi\nu^{(\pm)}\Gamma^2\left(-\nu^{(\pm)}\right)}\, ,\\
\tilde{b}_{12}(k)&=\frac{i\pi[ C^{(R)}(k)]^{-1}}{[\Gamma\left(-\nu^{(\pm)}\right)]^{-2}\sin\pi\nu^{(\pm)}}\, ,
\ \   \tilde{b}_{21}(k)= \frac{i}{\pi} \Gamma^2\left(1-\nu^{(\pm)}\right)\sin\pi\nu^{(\pm)} C^{(R)}(k)\, ,
\end{align}
\end{subequations}
with the $C^{(L)}(k)$  and $C^{(R)}(k)$ functions defined in (\ref{aa9}). Like in the space-like case we obtain
\begin{align}\label{Bmmtime}
B_{--}=&\frac{i\pi }{2\sin^2\pi \nu}\left(\frac{d^{(0)}(k_0)}{x^{1/2}}+\frac{V^{(-,0)}_{21}(-k_F)}{x}+\frac{V^{(+,0)}_{21}(k_F)}{x}\right)
\, , \ \ \ \nu=\frac{\kappa}{2}-\frac{\eta}{2\pi}\, ,\nonumber\\
=&-\frac{\pi e^{2i\pi\nu}}{2 \sin^2\pi \nu}\left(e^{-2i\pi\nu}-1\right)^2 \overline{G}(x,t)\left(\frac{2k_F t+x}{2k_Ft-x}\right)^{2\nu}\nonumber\\
 &\ \ \ +\frac{\pi^2\left(e^{-2i\pi\nu}-1\right)}{2\sin^3\pi\nu}\left(\frac{(2k_F)^{-2\nu}}{\Gamma^2(-\nu)}\frac{e^{i k_F x}}{(2k_F t+x)^{2\nu+1}}
+e^{2i\pi\nu} \frac{(2k_F)^{2\nu}}{\Gamma^2(\nu)}\frac{e^{-i k_F x}}{(2k_F t-x)^{-2\nu+1}}
\right)\, ,
\end{align}
and
\begin{align}\label{bpptime}
b_{++}=&-i\left(\frac{V^{(-,0)}_{12}(-k_F)}{x}+\frac{V^{(+,0)}_{12}(k_F)}{x}\right)\, ,\ \ \ \nu=-\frac{\kappa}{2}+\frac{\eta}{2}\, ,\nonumber\\
=& -\frac{\pi\left(e^{-2i\pi\nu}-1\right)^{-1}}{\sin\pi\nu}\left(\frac{(2k_F)^{2\nu}}{\Gamma^2(\nu)}\frac{e^{-i k_F x}}{(2k_F t+x)^{-2\nu+1}}
+e^{-2i\pi\nu}\frac{(2k_F)^{-2\nu}}{\Gamma^2(-\nu)}\frac{e^{i k_F x}}{(2k_F t-x)^{2\nu+1}}
\right)\, ,
\end{align}
where we have used $\Gamma(1-z)\Gamma(z)=\pi/\sin\pi z\, .$

\section{Thermodynamic limit of singular sums}\label{a5}

In this Appendix we are going to rewrite the elements of the $S^{(\pm)}$  and $R^{(+)}$ matrices in
a form that will allow to take the thermodynamic limit. We are going to need two identities (1.421(3) and 1.422(4) of \cite{GR}):
\begin{subequations}
\begin{align}
\cot \pi x&=\frac{1}{\pi x}+\frac{1}{\pi}\sum_{j=1}^\infty\left(\frac{1}{x-j}+\frac{1}{x+j}\right)\, ,\label{ident1}\\
\frac{1}{\sin^2\pi x}&=\frac{1}{\pi^2}\sum_{j=-\infty}^\infty\frac{1}{(x-j)^2}\, .\label{ident2}
\end{align}
\end{subequations}

\subsection{Thermodynamic limit of $S^{(-)}$}

The quasimomenta appearing in the definition of the $S^{(-)}$ matrix (\ref{defsminus}) can be written as
\be\label{difference}
k_j=\frac{2\pi}{L}j-\frac{\pi\kappa N}{L}+\frac{\Lambda}{L}\, ,\ \ q_l=\frac{2\pi}{L}l-\frac{\pi\kappa (N-1)}{L}+\frac{\Theta}{L}\, ,\ \ \  j,l\in\mathbb{Z}\, ,
\ee
which means that $q_l-k_j=2\pi m/L+\pi\kappa/L+(\Theta-\Lambda)/L$ with $m\in\mathbb{Z}$. Using this relation and the identities (\ref{ident1}) and (\ref{ident2})
we obtain two useful formulae $(q=\tilde{q}+\Theta/L)$
\be\label{identminus}
\frac{2}{L}\sum_{\tilde{q}}\frac{1}{q-k}=-\cot\left(\frac{\Lambda-\Theta-\pi\kappa}{2}\right)\, ,\ \
\frac{|1-e^{i\pi\kappa}\bar{\omega}\zeta|^2}{L^2}\sum_{\tilde{q}}\frac{1}{(q-k)^2}=1\, .
\ee
Using the first formula we can prove the following identity
\begin{align}\label{aa13}
\frac{|1-e^{i\pi\kappa}\bar{\omega}\zeta|^2}{L}\sum_{\tilde{q}}\frac{e^{-it q^2+iqx}}{q-k}&=
4\sin^2\left(\frac{\Lambda-\Theta-\pi\kappa}{2}\right)\frac{1}{L}\left\{\sum_{\tilde{q}} \frac{e^{-it q^2+iqx}-e^{-it k^2+ikx}}{q-k}
+\sum_{\tilde{q}} \frac{e^{-it k^2+ikx}}{q-k}\right\}\, ,\nonumber\\
&=\left[1-\cos\left(\Lambda-\Theta-\pi\kappa\right)\right]e^{(-)}(k)-\sin\left(\Lambda-\Theta-\pi\kappa\right)[e_-(k)]^{-2}\, ,
\end{align}
where we have introduced the function
\be\label{defeminusup}
e^{(-)}(k)=\frac{2}{L}\sum_{\tilde{q}} \frac{e^{-it q^2+iqx}-e^{-it k^2+ikx}}{q-k}
\ee
and $e_-(k)$ is defined in (\ref{defeminus}). Now we have all the necessary tools to rewrite the $S^{(-)}$ matrix. We will consider
first the off-diagonal elements $(a\ne b)$. Starting from the definition (\ref{defsminus}) and successively using the identities
$(k_a-q)^{-1}(k_b-q)^{-1}=(k_a-k_b)^{-1}[(q-k_a)^{-1}-(q-k_b)^{-1}]$ and (\ref{aa13}) we obtain
\begin{align}\label{aa14}
[S^{(-)}]_{ab}&=e_-(k_a)e_-(k_b)\frac{|1-e^{i\pi\kappa}\bar{\omega}\zeta|^2}{L^2(k_a-k_b)}
\sum_{\tilde{q}}\left(\frac{e^{-itq^2+iqx}}{q-k_a}-\frac{e^{-itq^2+iqx}}{q-k_b}\right)\, , \nonumber\\
&=\frac{2}{L}\left\{\frac{1-\cos(\Lambda-\Theta-\pi\kappa)}{2}\frac{e_+^{(-)}(k_a)e_-(k_b)-e_-(k_a)e_+^{(-)}(k_b)}{k_a-k_b}\right.\nonumber\\
&\left.\qquad\qquad\qquad -\frac{\sin(\Lambda-\Theta-\pi\kappa)}{2}\frac{[e_-(k_a)]^{-1}e_-(k_b)-e_-(k_a)[e_-(k_b)]^{-1}}{k_a-k_b}
\right\}\, ,
\end{align}
with $e_+^{(-)}(k)=e^{(-)}(k)e_-(k)$. What about the diagonal elements? Let us show that in the thermodynamic limit
 $[S^{(-)}]_{aa}=1+\lim_{k_a\rightarrow k_b}[S^{(-)}]_{ab}$ with $[S^{(-)}]_{ab}$ written as in (\ref{aa14}). Denoting $k_a$
 by $k$ in order to  lighten the notation the diagonal elements of the $S^(-)$ matrix (\ref{defsminus}) can be written as
\be\label{aa15}
[S^{(-)}]_{aa}=e_-(k)^2 \left[\frac{|1-e^{i\pi\kappa}\bar{\omega}\zeta|^2}{L^2}\sum_{\tilde{q}}\frac{e^{-it q^2+iqx}-e^{-it k^2+ikx}}{(q-k)^2}+e^{-it k^2+i k x}\right]\, ,
\ee
where we have used the second identity from (\ref{identminus}). From the definition (\ref{defeminusup}) we have
\be
[e^{(-)}(k)]'=\frac{2}{L}\sum_{\tilde{q}}\frac{e^{-it q^2+iqx}-e^{-it k^2+ikx}}{(q-k)^2}-\frac{2}{L}\sum_{\tilde{q}}\frac{(-2it k+ix)e^{-it k^2+ikx}}{q-k}\, ,
\ee
and using the first identity in (\ref{identminus}) it is easy to see that $e^{-itk^2+ikx}|1-e^{i\pi\kappa}\bar{\omega}\zeta|^2 \sum_{\tilde{q}} 2/[L(q-k)]
=-2\sin(\Lambda-\Theta-\pi\kappa)[e_-(k)]^{-2}.$ Plugging these intermediary results in (\ref{aa15}) we find
\be
[S^{(-)}]_{aa}=1+  \frac{1}{L}\left\{  (1-\cos(\Lambda-\Theta-\pi\kappa)[e_-(k)]^2[e^{(-)}(k)]'-\sin(\Lambda-\Theta-\pi\kappa)(-2itk+ix) \right\}\, ,
\ee
which is exactly what we would obtain by taking the limit $k_a\rightarrow k_b$ and using l'H\^{o}pital's rule in (\ref{aa14}).

\subsection{Thermodynamic limit of $S^{(+)}$ and $R^{(+)}$}

The functions appearing in the definitions (\ref{defsplus}), (\ref{defrplus}) of $S^{(+)}$ and $R^{(+)}$ involve sums over $\tilde{k}$'s. The general relation for
the difference of the quasimomenta (\ref{difference}) remains the same and the building identities become
\be\label{identplus}
\frac{2}{L}\sum_{\tilde{k}}\frac{1}{k-q}=\cot\left(\frac{\Lambda-\Theta-\pi\kappa}{2}\right)\, ,\ \
\frac{|1-e^{i\pi\kappa}\bar{\omega}\zeta|^2}{L^2}\sum_{\tilde{k}}\frac{1}{(k-q)^2}=1\, .
\ee
Introducing the function $e^{(+)}(q)$ defined by
\be
e^{(+)}(q)=\frac{2}{L}\sum_{\tilde{k}}\frac{e^{-i t k^2+i k x}-e^{-i t q^2+i q x}}{k-q}\, ,
\ee
and using the first identity in (\ref{identplus}) we obtain (note the sign change in front of the $\sin$ term compared with (\ref{aa13}))
\begin{align}\label{aa16}
\frac{|1-e^{i\pi\kappa}\bar{\omega}\zeta|^2}{L}\sum_{\tilde{k}}\frac{e^{-it k^2+ikx}}{k-q}
&=\left[1-\cos\left(\Lambda-\Theta-\pi\kappa\right)\right]e^{(+)}(k)+\sin\left(\Lambda-\Theta-\pi\kappa\right)[e_-(k)]^{-2}\, .
\end{align}
In a manner similar with the previous section we obtain for the off diagonal elements $(q_q\ne q_b)$ (again, note the sign change in
front of the $\sin$ term)
\begin{align}\label{aa17}
[S^{(+)}]_{ab}&=\frac{2}{L}\left\{\frac{1-\cos(\Lambda-\Theta-\pi\kappa)}{2}\frac{e_+^{(+)}(q_a)e_-(q_b)-e_-(q_a)e_+^{(+)}(q_b)}{q_a-q_b}\right.\nonumber\\
&\left.\qquad\qquad\qquad +\frac{\sin(\Lambda-\Theta-\pi\kappa)}{2}\frac{[e_-(q_a)]^{-1}e_-(q_b)-e_-(q_a)[e_-(q_b)]^{-1}}{q_a-q_b}
\right\}\, ,
\end{align}
with $e_+^{(+)}(k)=e^{(+)}(k)e_-(k)$. The diagonal terms are computed using l'H\^{o}pital's rule. In the case of $R^{(+)}$ using the identity (\ref{aa16}) we find
\begin{align}
[R^{(+)}]_{ab}&=\frac{1-\cos(\Lambda-\Theta-\pi\kappa)}{2}\frac{e^{(+)}_+(q_a)e^{(+)}_+(q_b)}{L}
+\frac{\sin(\Lambda-\Theta-\pi\kappa)}{2}\left[\frac{e_+^{(+)}( q_a)}{L e_-(q_b)}+\frac{e_+^{(+)}(q_b)}{L e_-(q_a)}\right]\nonumber\\
&\qquad\qquad\qquad\qquad\qquad +\frac{1+\cos(\Lambda-\Theta-\pi\kappa)}{2}\frac{1}{Le_-(q_a)e_-(q_b)}\, .
\end{align}

\twocolumngrid

\end{document}